\documentclass[pdflatex,sn-mathphys-num]{sn-jnl}

\usepackage{graphicx}%
\usepackage{multirow}%
\usepackage{amsmath,amssymb,amsfonts}%
\usepackage{amsthm}%
\usepackage{mathrsfs}%
\usepackage[title]{appendix}%
\usepackage{xcolor}%
\usepackage{textcomp}%
\usepackage{manyfoot}%
\usepackage{booktabs}%
\usepackage{algorithm}%
\usepackage{algorithmicx}%
\usepackage{algpseudocode}%
\usepackage{listings}%
\usepackage{aas_macros}

\usepackage{siunitx}
\usepackage{ulem}

\usepackage{etoolbox}

\usepackage{caption}
\usepackage{multibib}
\newcites{supp}{  }

\theoremstyle{thmstyleone}%
\theoremstyle{thmstyletwo}%

\theoremstyle{thmstylethree}%

\newcommand{\LIRA}		 	{1}
\newcommand{\ESOGarching}	{2}
\newcommand{\FEUP}		 	{3}
\newcommand{\CENTRA}	 	{4}
\newcommand{\Valencia}    {5}
\newcommand{\MPE}		 	{6}
\newcommand{\IPAG}		 	{7}
\newcommand{\MPIA}		 	{8}
\newcommand{\LMU}	 	 	{9}
\newcommand{\MPA} {10}
\newcommand{\Vina} {11}
\newcommand{\TITANS} {12}
\newcommand{\KUL}	{13}
\newcommand{\TUM}	 	 	{14}
\newcommand{\UCD}		{15}
\newcommand{\AUSTIN}    {16}
\newcommand{\UBerkley}   	{17}
\newcommand{\ESOSantiago}	{18}
\newcommand{\USouthampton}	{19}
\newcommand{\FCLA} {20}
\newcommand{\UCologne}	 	{21}
\newcommand{\CotedAzur}  {22}
\newcommand{\Technion}  {23}
\newcommand{\Hebrew}    {24}
\newcommand{\UNAM}		{25}
\newcommand{\CRAL}		{26}

\begin{document}

\title[Discovery of a star sensitive to the spin of Sgr~A*]{Discovery of a star sensitive to the spin of Sgr~A*}

\author{
	\begin{center}
		\linespread{1.2}\selectfont
		K.~Abd El Dayem$^{\LIRA}$,
        R.~Abuter$^{\ESOGarching}$,         
		N.~Aimar$^{\FEUP,\CENTRA}$,  
        P.~Amaro-Seoane$^{\Valencia,\MPE}$,
		A.~Berdeu$^{\ESOGarching,\LIRA}$,
		J.-P.~Berger$^{\IPAG}$,      
		G.~Bourdarot$^{\MPE}$,
		W.~Brandner$^{\MPIA}$,    
        A.~Burkert$^{\LMU,\MPE}$,
        D.~Calderon$^{\MPA}$,
		C.~Correia$^{\FEUP,\CENTRA}$,
        J.~Cuadra$^{\Vina, \TITANS}$,
		R.~Davies$^{\MPE}$,
		D.~Defr{\`e}re$^{\KUL}$,
        L.~Delit$^{\LIRA}$,
		A.~Drescher$^{\IPAG,\MPE}$,                         
		F.~Eisenhauer$^{\MPE,\TUM}$,
        L.~Esteras~Otal$^{\ESOGarching}$,
		M.~Fabricius$^{\MPE}$,
		H.~Feuchtgruber$^{\MPE}$,       
		N.M.~F{\"o}rster~Schreiber$^{\MPE}$, 
		A.~Foschi$^{\LIRA}$,        
		P.~Garcia$^{\FEUP,\CENTRA}$,        
		R.~Garcia~Lopez$^{\UCD}$,
        A.~Generozov$^{\AUSTIN}$,
		R.~Genzel$^{\MPE,\UBerkley}$,        
		S.~Gillessen$^{\MPE, \dagger}$,           
		F.~Gont{\'e}$^{\ESOGarching}$,
		X.~Haubois$^{\ESOSantiago}$,     
		S.F.~H{\"o}nig$^{\USouthampton}$,      
		M.~Houll{\'e}$^{\IPAG}$,
		S.~Joharle$^\MPE$,
		A.~Kaufer$^{\ESOSantiago}$,
        J.~Kammerer$^{\ESOGarching}$,
		P.~Kervella$^{\LIRA,\FCLA}$,   
        J.~Kolb$^{\ESOGarching}$,
		L.~Kreidberg$^{\MPIA}$,
		L.~Labadie$^{\UCologne}$,
		S.~Lacour$^{\LIRA}$,
		O.~Lai$^{\CotedAzur}$,
		R.~Laugier$^{\KUL}$,
		J.-B.~Le~Bouquin$^{\IPAG}$,
		J.~Leftley$^{\ESOSantiago,\USouthampton}$,
		B.~Lopez$^{\CotedAzur}$,
		D.~Lutz$^{\MPE}$,           
		F.~Mang$^{\MPE,\TUM,\dagger}$,
		A.~M{\'e}rand$^{\ESOGarching}$,
		F.~Millour$^{\CotedAzur}$,
		M.~Montarg{\`e}s$^{\LIRA}$,           
		N.~Moruj{\~a}o$^{\FEUP,\CENTRA}$, 
		H.~Nowacki$^{\CotedAzur}$, 
		M.~Nowak${^\LIRA}$, 
        S.~Oberti${^\ESOGarching}$,
		J.~Osorno$^{\LIRA,\dagger}$,
		T.~Ott$^{\MPE}$,           
		T.~Paumard$^{\LIRA}$,
        C.~Paladini$^{\ESOSantiago}$,
        H.B.~Perets$^{\Technion}$,
		K.~Perraut$^{\IPAG}$, 
		G.~Perrin$^{\LIRA}$,
		R.~Petrov$^{\CotedAzur}$,
		P.O.~Petrucci$^{\IPAG}$,  
        T.~Piran$^{\Hebrew}$,
		N.~Pourr{\'e}$^{\IPAG}$,           
		S.~Rabien$^{\MPE}$,
		D.C.~Ribeiro$^{\MPE}$,           
		S.~Robbe-Dubois$^{\CotedAzur}$,
		M.~Sadun~Bordoni$^{\MPE}$,
		J.~S{\'a}nchez~Berm{\'u}dez$^{\UNAM}$,
		D.~Santos$^{\MPE}$, 
        R.~Sari$^{\Hebrew}$,
		J.~Sauter$^{\MPIA}$,
        S.~Scheithauer$^{\MPIA}$,
		J.~Scigliuto$^{\CotedAzur}$,  
		J.~Shangguan$^{\MPE}$,   
		T.T.~Shimizu$^{\MPE}$,   
		F.~Soulez$^{\CRAL}$,
        J.~Stadler$^{\LMU}$,
		C.~Straubmeier$^{\UCologne}$,
		E.~Sturm$^{\MPE}$,   
		M.~Subroweit$^{\UCologne}$,   
		C.~Sykes$^{\USouthampton}$,
		L.J.~Tacconi$^{\MPE}$,   
        P.~Th{\'e}venet$^{\LIRA}$,
        I.~Urso$^{\LIRA}$,
		F.~Vincent$^{\LIRA}$,   
		J.~Woillez$^{\ESOGarching}$,
        G.~Zins$^{\ESOSantiago}$ 
    
	\end{center}

	\begin{center}
		\linespread{1.2}\selectfont
		\small
		
		$^{\LIRA}$
		\textit{LIRA, Observatoire de Paris, Universit{\'e} PSL, CNRS, Sorbonne Universit{\' e}, Universit{\' e} de Paris, 5 place Jules Janssen, 92195 Meudon, France   }

		$^{\ESOGarching}$
		\textit{European Southern Observatory, Karl-Schwarzschild-Stra{\ss}e 2, 85748 Garching, Germany }

		$^{\FEUP}$
		\textit{Faculdade de Engenharia, Universidade do Porto, rua Dr. Roberto Frias, 4200-465 Porto, Portugal }

		$^{\CENTRA}$
		\textit{CENTRA - Centro de Astrof{\' i}sica e Gravita\c{c}{\~a}o, IST, Universidade de Lisboa, 1049-001 Lisboa, Portugal }

	        $^{\Valencia}$
       		\textit{Universitat Polit{\`e}cnica de Val{\`e}ncia, C/Vera s/n, Val{\`e}ncia, 46022, Spain}

		$^{\MPE}$ 
		\textit{Max Planck Institute for Extraterrestrial Physics, Giessenbachstra{\ss}e 1, 85748 Garching, Germany }

		$^\IPAG$
		\textit{Univ. Grenoble Alpes, CNRS, IPAG, 38000 Grenoble, France }

		$^\MPIA$
		\textit{Max Planck Institute for Astronomy, K{\"o}nigstuhl 17, 69117 Heidelberg, Germany}

		$^{\LMU}$
		\textit{University Observatory, Faculty of Physics, Ludwig-Maximilians-Universit{\"a}t, Scheinerstra{\ss}e 1, 81679 Munich, Germany}
	
        		$^{\MPA}$ 
		\textit{Max Planck Institute for Astrophysics, Karl-Schwarzschild-Stra{\ss}e 1, 85748 Garching, Germany }

	        $^{\Vina}$
       		 \textit{Universidad Adolfo Ib{\'a}\~nez, Av. Padre Hurtado 750, Vi\~na del Mar, Chile }

 	       	$^{\TITANS}$
       		\textit{Millennium Nucleus on Transversal Research and Technology to Explore Supermassive Black Holes (TITANS), Chile}

		$^{\KUL}$
		\textit{Institute of Astronomy, KU Leuven, Celestijnenlaan 200D, 3001, Leuven, Belgium}

		$^{\TUM}$
		\textit{Department of Physics, TUM School of Natural Sciences, Technical University of Munich, 85748 Garching, Germany }

		$^{\UCD}$
		\textit{School of Physics, University College Dublin, Belfield, Dublin 4, Ireland}

       	 	$^{\AUSTIN}$
      		 \textit{Astronomy Dept. and Oden Institute, University of Texas at Austin, Austin, TX 78712, USA}

		$^{\UBerkley}$
		\textit{Departments of Physics \& Astronomy, Le Conte Hall, University of California, Berkeley, CA 94720, USA }

		$^{\ESOSantiago}$
		\textit{European Southern Observatory, Casilla 19001, Santiago 19, Chile }

		$^{\USouthampton}$
		\textit{School of Physics \& Astronomy, University of Southampton, Southampton, SO17 1BJ, United Kingdom}

		 $^{\FCLA}$
		\textit{French-Chilean Laboratory for Astronomy, IRL 3386, CNRS and U. de Chile, Casilla 36-D, Santiago, Chile.}
		
		$^{\CotedAzur}$
		\textit{Universit{\'e} C{\^o}te d{'}Azur, Observatoire de la C{\^o}te  d{'}Azur, CNRS, Laboratoire Lagrange, France}

    		 $^{\Technion}$
        		\textit{Physics department, Technion - Israel Institute of Technology, Technion city, Haifa 3200002, Israel}

     		 $^{\Hebrew}$
        		\textit{Racah Institute of Physics, The Hebrew University, Jerusalem 91904, Israel}

		$^{\UNAM}$
		\textit{Instituto de Astronom{\'i}a, National Autonomous University of Mexico, Mexico City, Mexico}

		$^{\CRAL}$
		\textit{Univ. Lyon, Univ. Lyon 1, ENS de Lyon, CNRS, Centre de Recherche Astrophysique de Lyon UMR5574, F-69230, Saint Genis-Laval, France}

		$^{\UCologne}$
		\textit{1st Institute of Physics, University of Cologne, Z{\"u}lpicher Stra{\ss}e 77, 50937 Cologne, Germany }
		
		$^\dagger$
		Corresponding authors: F.~Mang (fmang@mpe.mpg.de), J.~Osorno (juan.osorno@obspm.fr), S.~Gillessen (ste@mpe.mpg.de).

	\end{center}
}

\abstract{Residing in the center of the Milky Way, Sgr~A* is the closest massive black hole (MBH, \citep{2021arXiv210213000G}). Its vicinity has allowed measuring individual stellar orbits around it \citep{schodel_star_2002,2008ApJ...689.1044G,gillessen_monitoring_2009}. The stars act as test particles and probe the gravitational potential around the $4.3\times 10^6 M_\odot$ MBH. These observations have determined the central mass to sub-percent precision \citep{gravitycollaborationMassDistributionGalactic2022}, and the mildly relativistic motions of stars have given access to  the dominant relativistic corrections, the gravitational redshift \citep{2018A&A...615L..15G, 2019Sci...365..664D}, the transverse Doppler effect, and the prograde precession imposed by the Schwarzschild metric nature of the potential \citep{2020A&A...636L...5G}. These effects are of order $\beta^2=(v/c)^2$ (for velocity $v$ and speed of light $c$). The Kerr metric for a rotating black hole leads to corrections of order $\beta^3$.
Here, we report the discovery of a faint main-sequence star ($m_K = 19.3)$, S301, on a 8.7-year orbit and with small enough a pericenter distance, such that the star's peak velocity reaches $25000\,$km/s. Within the measurement capabilities of current near-infrared interferometry and future spectroscopy on an extremely large telescope, 
S301's motion is directly sensitive to the spin of Sgr~A*. The high eccentricity of S301 suggests that it is the captured component of a binary that was torn apart via the Hills mechanism.}


\maketitle
Black holes in general relativity have just two additional degrees of freedom beyond mass: spin and charge. Astrophysically relevant is the spin. As all objects in the Universe rotate, one  expects the same for black holes, in particular, since the angular momentum of material creating a black hole is conserved. Since the effects of the spin on space-time fall off with distance $r$ to the black hole like $r^{-3}$, it is actually hard to measure a spin.  The existence of jets in active galactic nuclei requires that the MBHs located in the central engines rotate \cite{2019ARA&A..57..467B}.  Spin estimates have been obtained from X-ray reflection spectra, in which the iron K-$\alpha$ line shape is a probe of the spin, albeit the spin's impact is small \citep{1995Natur.375..659T}. For accreting stellar-mass black holes, spins can be estimated from accretion theory \citep{McClintock+2006}, and thus are not assumption-free. The cleanest signatures are probably those of gravitational wave mergers, in which the spins of the initial objects are among the fit parameters to the pre-mergers wave forms~\citep{2016PhRvL.116x1103A} and the spin of the resulting black hole can be inferred from its ringdown signature~\citep{LVK2025_GW250114}. The space experiment Gravity Probe B detected the spin-induced precession due to Earth twisting space-time with 5$\sigma$ significance~\citep{2011PhRvL.106v1101E}, testing the far-field and slow-motion approximation around a massive body, which equals the
approximation of the Kerr metric in the same limit \citep{Adler_2015}.
Overall, there are only few  observational constraints on the spin parameter of the Kerr metric. 

\noindent Sgr~A*, the closest MBH in the Galactic Center (GC) at a  distance of $8.3\,$kpc offers a direct, dynamical way to measure its spin. The observation of stellar orbits has made Sgr~A* one of the best cases for the existence of black holes in general. A few dozen stars revolve on (nearly) Keplerian orbits, with an almost relaxed eccentricity distribution and randomly oriented orbits.  Most valuable are the stars that come closest to Sgr~A*, as they probe deepest into the gravitational potential. In particular the star S2 on a 16-year orbit \citep{schodel_star_2002} has been in focus, due to its comparably easily accessible orbit. S2's motion is notably affected by relativistic effects: During its 2018 pericenter passage, the gravitational redshift of Sgr~A* led to an additional change of measured atomic line positions of  $\approx 200\,$km/s \citep{2018A&A...615L..15G, 2019Sci...365..664D}. By 2020, the astrometric data of the star showed that the orbit had precessed in 2018 by around 12', fully consistent with  a motion in the Schwarzschild metric \citep{2020A&A...636L...5G}. Key for these discoveries was the advent of near-infrared interferometry with the GRAVITY instrument  at the European Southern Observatory's Very Large Telescope (VLT) \citep{2017A&A...602A..94G}. At the $\beta^3$ order, the leading-order term from the Kerr metric contributes, and the motion is sensitive to the spin of Sgr~A*, often called `Lense-Thirring' precession \citep{Lense:1918zz, 2005ApJ...622..878W, 2010ApJ...720.1303A, 2016ApJ...818..121P}.

\noindent For S2, current instrumentation  requires prohibitively long time series to detect the spin. The parameter determining how sensitive a star is to relativistic effects is the pericenter distance $r_p = a(1-e)$ \citep{2018MNRAS.476.3600W}, which for S2 is $1400\,R_S$ (Schwarzschild radii, $1 R_S = 10\,\mu$as for Sgr~A*). Stars with smaller $r_p$ would allow a quicker detection, requiring smaller semi-major axes $a$ and/or larger eccentricities $e$. Given a spatial resolution of $\approx 1.7\,$mas, an astrometric precision of $\simeq\SI{30}{\mu as}$ comparable to $R_S$  and a field of view of  $\approx 70\,$mas, GRAVITY can trace stars with pericenter passages much closer to Sgr~A* than S2. The newly discovered star S301 constitutes  a first example for such a relativistic test particle suitable for measuring the spin of Sgr~A*.

\section*{Observations \& data analysis}
Since 2017, we have regularly observed the central arcsecond around Sgr~A* with GRAVITY  in order to track the stellar motions. Observations take place monthly during roughly week-long campaigns between March and September (Extended Data Table~1). A total of around 80-100 hours per year of observing time is used for that. A major part of it is spent on a central pointing containing Sgr~A*, as this serves as astrometric reference for all stellar positions. The data consist per $360\,$s exposure of a set of complex visibilities for six baselines, sampled at 12 spectral channels for the two linear polarizations (Extended Data Fig.~1).

\noindent We group the data per night and analyze the data sets in two ways: First, we fit a model to them. It contains the known sources and returns positions and fluxes of these. By design, this does not reveal new sources. In a second step, we employ image reconstruction  to Fourier-invert the data into an image. Beyond classical algorithms developed for radio interferometry like CLEAN (for an example see Extended Data Fig.~2), we use our own code `GRAVITY-RESOLVE', $G^R$ (\citep{gravity_collaboration_deep_2022,mang_in_prep}, Methods section). The images  allow identifying previously unknown sources (Fig.~1).

\noindent In spring 2023, we discovered a faint star $15\,$mas north-west of Sgr~A*, which in the following months  moved outward, and which we labeled S301. From the four positions in 2023 we derived a fast  ($v  \approx (-44,+27)\, \mathrm{mas} \, \mathrm{yr}^{-1}, |v|  \approx 2000\,\mathrm{km} \,\mathrm{s}^{-1}$) and slightly curved motion  (significant at the $+3 / +5 \,\sigma$ level), that would bring the star by 2024 outside of the central field of view.
We followed S301 with dedicated pointings in 2024 and 2025, yielding eight and five additional astrometric measurements  respectively. 
Given the preliminary orbit, we were  able to post-dict  the star's positions  in previous observing epochs (Extended Data Fig.~3), and whether it would be by chance in one of our pointings. We found a strong inference of S301 in 2021 and a weak inference in 2017 (Extended Data Fig.~4). These positions are less precise, as only few exposures at the respective pointings were taken. Overall, we have  19 astrometric positions of S301 that outline an ellipse  on the sky and yield a consistent orbit.

\noindent We have tried to identify S301 in deep ERIS integral-field spectroscopic data, but have not been able to detect it as a continuum source or from spectral features.
We thus do not yet have any radial velocity information.

\begin{figure}[h]
    \centering
    \includegraphics[width=0.85\linewidth]{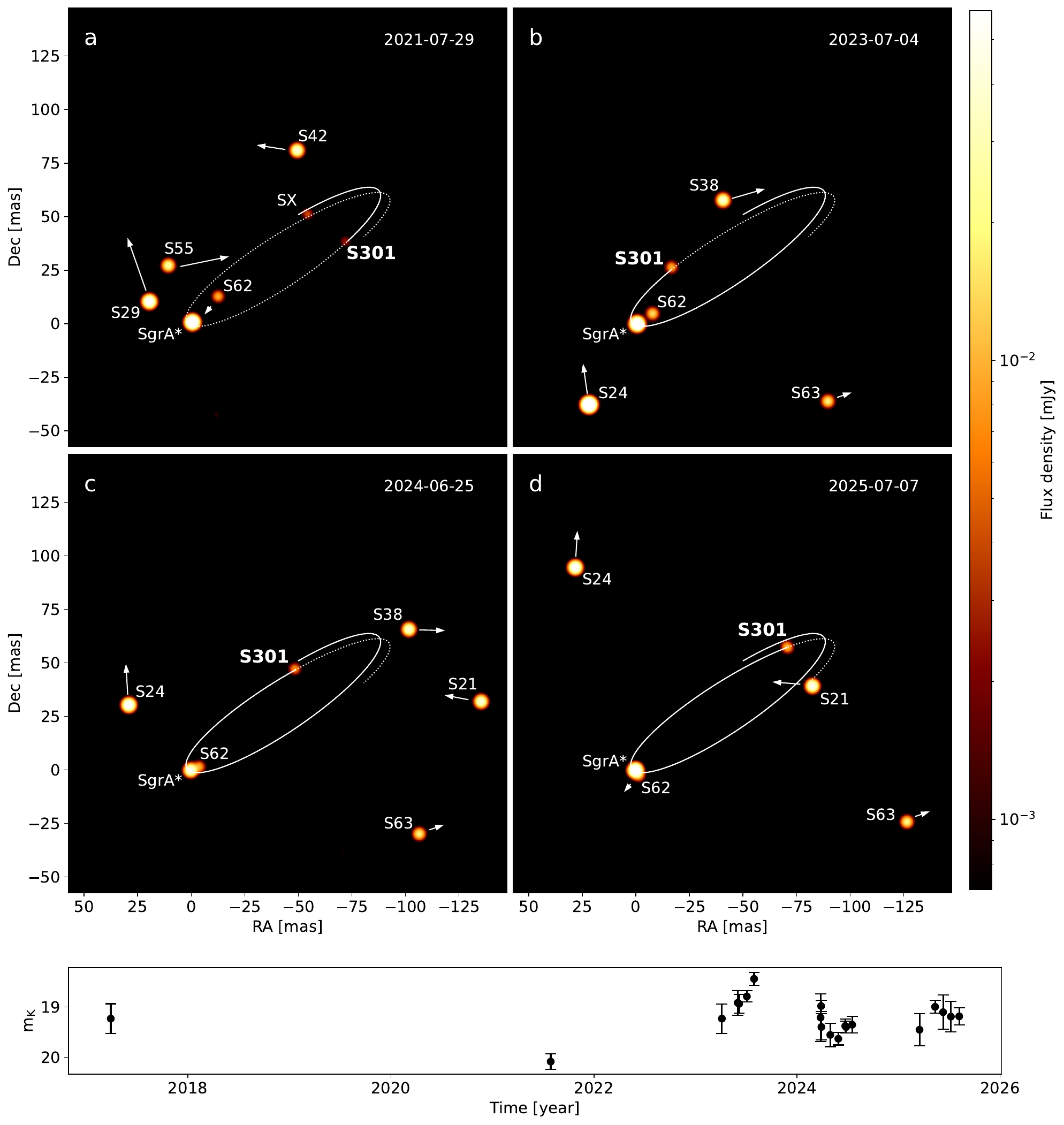}
 \caption{{\bf GRAVITY images of the region around Sgr~A*.} Top: Annotated time series from 2021 to 2025, reconstructed with $\text{G}^{\text{R}}$.
 For S301 the orbital trace (Fig.~2) is shown on top. In panels \textbf{c} and \textbf{d}, Sgr~A* and S62 are confused  in this representation due to the smoothing of the image with a Gaussian kernel. SX in the 2021 image is a potential source not yet rediscovered.   Bottom: K-band light curve of S301.}
\end{figure}

\section*{Results}
From the GRAVITY images, we measure the positions of S301 and Sgr~A* to obtain the star's position vector relative to the MBH. For observations, in which the S301 pointing was not centered on Sgr~A* (in 2017, 2021, 2024 and 2025), the two objects are fitted relative to their respective field centers. The positional offset between the latter two is measured at the interferometric precision by GRAVITY's metrology system. Our imaging code $G^R$ also allows estimating the uncertainties in the positions, as multiple instances of the same images are inferred.
We measure the brightness of the objects using the brighter sources in the images, for which photometry is available from \cite{gillessen_monitoring_2009}. For S301, we derive a K-band magnitude of $m_K = 19.3 \pm 0.3$. 

\noindent Using a  $\chi^2$-minimization, we fit a preliminary Keplerian orbit to the astrometric data ($\alpha(t)$, $\delta(t)$). According to this fit, S301 passed the pericenter of its orbit early 2023, with a 3D separation significantly smaller than that of S2. Hence, we need to take into account the $\beta^2$ relativistic effects, using the same model as for S2 in \cite{2020A&A...636L...5G}, including  the R{\o}mer effect (retardation effects due to the finite speed of light) and Schwarzschild precession. The potential is defined by the central mass fixed with $M_\mathrm{MBH} = 4.297\times 10^6 M_\odot$ at a distance of $R_0 = 8277\,$pc. We do not allow for any coordinate system offsets, as our data are interferometrically referenced directly to the near-infrared counterpart of Sgr~A*. 

\noindent Due to the lack of radial velocity information, two equally valid solutions exist, corresponding to the two possible orientations of the orbit. In principle, the R{\o}mer delay could break this degeneracy \citep{abuter_geometric_2019}, but the two  orientations yield indistinguishable best-fit $\chi^2$ values. Except for this sign ambiguity, the orbit fit converges uniquely (as verified by sampling the posterior space with a Markov chain, Extended Data Fig.~5) at $\chi^2=26$ for 34 degrees of freedom. The orbit does not agree with any previously claimed detections (Methods section).

\noindent The best-fit orbit is remarkable (Extended Data Table~2) with a semi-major axis of $a=83\,$mas (33\% smaller than S2's) and an orbital period of $8.7\,$years (Fig.~2). The star sets thus a new record for the shortest known orbital period around Sgr~A* (Extended Data Fig.~6 and~7), with S55 / S0-102 on a 12-year orbit \citep{2012Sci...338...84M} being the previous record-holder. Even more extreme is the eccentricity of $e=0.9832 / 0.9821$ (for the two possible  orientations), leading to a pericenter distance $r_p$ of only $136 / 142\,R_S$,  around ten times smaller than S2's. This implies  correspondingly stronger relativistic effects. The relativistic pericenter advance per orbit amounts to $2.0^\circ / 1.9^\circ$, such that after just 1560 / 1630 years the orbit in its plane has revolved once. At pericenter, S301 moves with $\approx 25\,600\,$ / $25\,000\,$km~s$^{-1}$ or 8.5\% / 8.3\% of $c$.

\begin{figure}[h]
    \centering
    \includegraphics[width=0.82\linewidth]{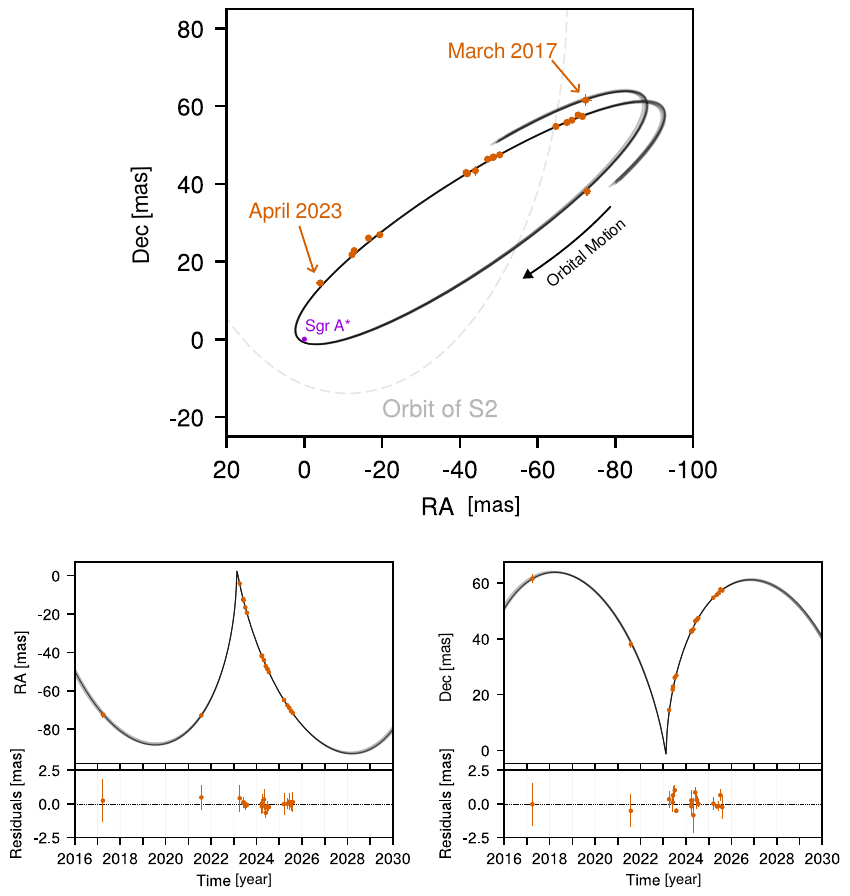}
    \caption{{\bf Orbit of S301 from 2016 to 2030.}
The upper panel shows the best-fit relativistic orbit to the positions in a Schwarzschild metric around  Sgr~A* (Methods section). The bottom panels show the motion and residuals over time separately in RA and Dec.  Due to the high eccentricity and short semi-major axis the orbit visibly does not close as a result of the Schwarzschild precession.}
\end{figure}

\section*{Discussion}
\subsection*{Stellar type}
With $m_K=19.3 \pm 0.3$ S301 is  too faint to be a giant in the GC, but the magnitude matches a main-sequence star of spectral type  F1.5, corresponding to a mass $\lesssim 1.5\,M_\odot$ (Methods section).
Estimating the mass  from stellar evolutionary tracks yields a consistent value (Extended Data Fig.~8).
S301's radius is between $\sim 1.4\,R_\odot$ and $\sim1.6\,R_{\odot}$. The main-sequence life time of early F-type stars is around $2\times 10^9\,$yr.
This interpretation is consistent with the fact that S301  did not get tidally disrupted at pericenter.  A giant star with $r_\ast=0.5\,$AU and $m_\ast = 3\,M_\odot$ 
begins to lose a fraction of its envelope at its tidal radius ${r_t = r_\ast (m_\ast / M_\mathrm{MBH})^{1/3}}  \approx 500\,R_S > r_p$. The loss of  mass would affect the orbit at pericenter, which is not observed.  On the other hand, for a main-sequence star $r_t \approx 10\,R_S< r_P$, such that  it is safe from tidal disruption (and tidal heating, Methods section) .

\noindent Spectrally, we therefore expect S301 to show exclusively the Brackett-$\gamma$ absorption line, but no further features in the infrared K-band \citep{1996ApJS..107..281H}. Our current ERIS spectroscopy is not deep enough to test this. Future MICADO \citep{2021Msngr.182...17D} spectroscopy at the ELT will have no problem in seeing the spectral feature(s) and will be able to determine S301's radial velocity.

\subsection*{Spin sensitivity}
The most exciting aspect of S301's orbit is its extremely small pericenter distance, such that it is sensitive to the spin of Sgr~A* to a degree accessible within around a decade, using existing and/or planned instrumentation. Fig.~3 (left) predicts for a  maximally rotating 
black hole  ($\chi=1$; with $\chi \equiv c J / (G M_\mathrm{MBH}^2)$  the dimensionless spin parameter) the in-plane precession effects. The dominant effect is the Schwarzschild precession, advancing the pericenter by $1.9^\circ$ per revolution.  The (in-plane contribution of) the Lense-Thirring precession 
is $0.11^\circ \chi \cos \xi$ 
per revolution (where $\xi$ is the inclination between spin axis and orbital angular momentum, Methods section), comparable to the angle S2's orbit advances due to the Schwarzschild metric, and which we have measured  in \cite{2020A&A...636L...5G}. The time scale for the spin precession~\cite{2003ApJ...590L..33L} is $\approx 58000\,$years.
Forecasting future observations allows testing by when the data would be measurably sensitive to the spin. Using reasonable assumptions, we estimate that within a decade there is a fair chance to directly measure Sgr~A*'s spin from the S301 orbit (Fig.~3 right, Methods section, Extended Data Fig.~9).

\begin{figure}[ht]
    \centering
    \includegraphics[width=0.95\linewidth]{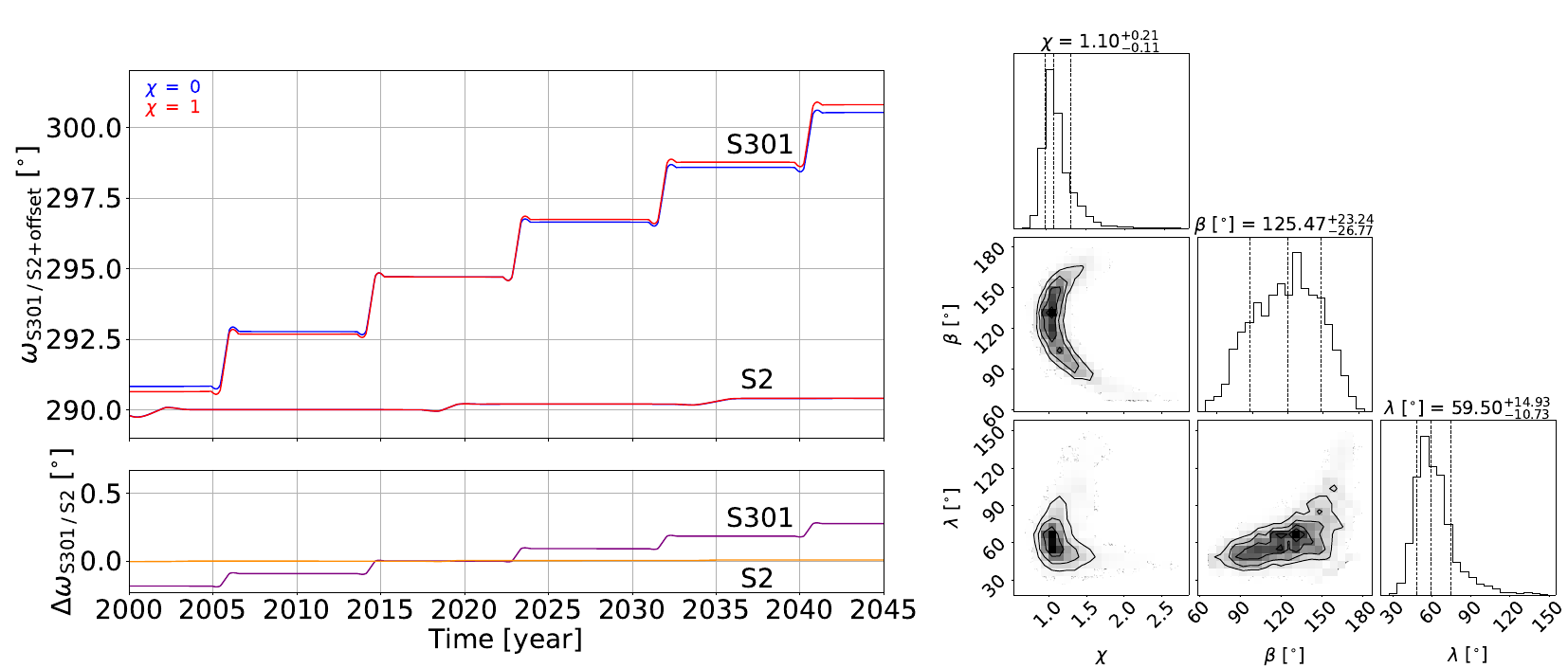}
    \caption{ {\bf Sensitivity of the S301 orbit to the spin of Sgr~A*.} 
Top left: Change of the orientation of the orbital ellipse (measured by the osculating longitude of pericenter $\omega$) as a function of time for S301 in comparison with S2. The discrete steps occur when the respective star passes the pericenter of its orbit. The blue lines are for a non-rotating MBH ($\chi = 0$), the red ones correspond to a ``randomly'' oriented ($\lambda=\xi=0$, pointing towards the Sun), maximally spinning ($\chi=1$) black hole. The extra-precession due to the spin of Sgr~A* of the S301 orbit is comparable to  the changes induced by the 1PN-terms of the Schwarzschild metric of S2, which GRAVITY  has detected already \citep{2020A&A...636L...5G, gravitycollaborationMassDistributionGalactic2022}.  Bottom left: the difference in $\omega$ between the two spin values. Right: Example of constraints on the spin parameters $(\chi,\lambda,\xi)$ for a mock data set with future simulated data up to 2035. These mock data  constrain  $\chi$ with an uncertainty $<0.2$, and also the orientation to around $\pm 30^\circ$. }
  \end{figure}

\noindent S301  provides the first and currently only practical way to measure the spin of Sgr~A* with stellar dynamics. While all other known S-stars owed to 
their larger pericenter distances are  insensitive to frame dragging on observationally accessible time scales, they will provide constraints on $M_{\rm MBH}$ and $R_0$, which can be used as  priors, reducing degeneracies and assisting the spin measurement. 
These stars would also constrain Newtonian perturbations due to dark masses orbiting Sgr A*, which are expected to be sub-dominant.
In the longer term, S301 opens a path towards constraining the quadrupole moment of Sgr~A* and testing the Kerr no-hair relation \citep{2008ApJ...674L..25W} once sufficiently long time series of sufficiently accurate astrometry and spectroscopy become available. 

\noindent S301's Schwarzschild precession offers the opportunity to tighten constraints on possible deviations from general relativity (e.g., strong-field PPN parameters) and on a putative extended mass distribution. The current dataset is consistent with general relativity at the 1PN level, although the resulting PPN bounds are weaker than those derived from S2  \citep{2020A&A...636L...5G, gravity_collaboration_mass_2024} with $f_\mathrm{SP} = 0.94 \pm 0.88$. A joint analysis may  improve these constraints, and continued monitoring of S301 will significantly strengthen them.

\subsection*{Origin}
Another interesting aspect of the orbit is its history, which we can constrain from the dynamical timescales (Extended Data Table~3, Methods section). One sees that the star and its orbit can be affected by stellar collisions and relaxation. 
Since star formation so close to the MBH is unlikely given the strong tidal effects, S301 has likely formed elsewhere and migrated to its current location. Given its  mass, S301's life time  is comparable to the two-body relaxation time, and it is therefore unlikely to have formed in the nuclear cluster and slowly migrated inwards through two-body relaxation. Instead, the extreme value $e\simeq0.98$ is naturally produced in  a Hills disruption of a compact main-sequence binary by the MBH (\citep{hillsHypervelocityTidalStars1988}, Methods section).

\noindent In the Hills picture, the timescale hierarchy at S301's radius, constrains its origin and evolution further. 
The angular momentum relaxation and collision time scales are shorter than the main-sequence lifetime. If the star had been in the GC for $ \gtrsim \mathrm{few} \times 10^7$ yr, its eccentricity would have thermalized, losing memory of its initial condition. In this case, the high eccentricity simply occurs by chance (there is a $\sim 3\%$ chance to find such a high eccentricity for a thermal distribution).
Alternatively, S301 may have been injected into the GC recently, so that it has not had time to relax, and constitutes a pristine fossil of its capture event.

\noindent The long two-body relaxation and gravitational wave inspiral times imply that the star's semi-major axis has not evolved significantly over its lifetime, and thus its present-day orbit constrains the progenitor binary. The measured semi-major axis is related to the one of the pre-disruption binary, yielding $a_\mathrm{bin}\sim 0.1~{\rm AU}$ (Methods section). Such compact binaries are common among F-type stars \citep{raghavan_binary_2010} and are expected to be tidally circularised and nearly synchronised. S301  would then have  an equatorial rotation velocity $v_\mathrm{rot}\sim 20$--$70~{\rm km\,s^{-1}}$, which  might be observable with future 
high-resolution spectroscopy from ELT/MICADO, providing a direct test of the Hills origin.

\noindent Taken together, the properties of S301 suggest a simple and self-consistent picture: a compact main-sequence binary was tidally separated by Sgr~A*, leaving behind S301 on the most relativistic stellar orbit known and ejecting its companion as a hyper-velocity star.

\bibliographystyle{plain}
\bibliography{sn-bibliography}


\newpage

\setcounter{figure}{0}
\renewcommand{\figurename}{Extended Data Fig.}
\renewcommand{\tablename}{Extended Data Table}

\section*{Methods}

\subsection*{Details of observations}
Our observations, from which we inferred S301, comprise 19 datasets spanning more than eight years. We first noted S301 in 2023, when pointing directly to Sgr~A*. Given the star's position close to Sgr~A* and its large proper motion, it was quickly clear that it might be on a tight orbit, and we 
followed up with dedicated observing campaigns over the course of the next years. We summarize the data used for this work in Extended Data Table~1.

\begin{table}[ht]
\caption{Summary of GRAVITY data that yielded an inference of S301.
The date in the first column follows ESO convention and gives the date of the evening of the respective  observing night. The epoch is the mean over the time stamps of the $\text{N}_{\text{exp}}$ individual GRAVITY exposures. The total accumulated integration time per night is given by $t_{\text{int}}$, and $\mathbf{p}$ is the center  of the individual pointings with respect to Sgr~A*. All data were acquired in low-resolution mode and in two linear polarizations, except for the 2017 dataset, which was partially acquired with both polarizations combined.}
\label{tab_dedicated_obs}
    \centering
    \begin{tabular}{c|c|c|c|c|c}
        Date & Epoch & $\text{N}_{\text{exp}}$ & $t_{\text{int}} \left[\text{h}\right]$ & $\mathbf{p} \left[\text{mas}\right]$ & program ID \\\hline
       2017-03-26..30 & 2017.2389 & 12 (one pol.) & 1.0 & $(-57.974, 35.939)$&60.A-9102(A)\\\hline
        2021-07-29 & 2021.5759 & 8 & 0.71 & $(-45., 45.)$ & 105.20B2.004\\\hline
        2023-04-05 & 2023.2604 & 12 & 1.2 & $(0.0, 0.0)$ & 111.24H1.001\\
        2023-06-01 & 2023.4162 & 13 & 1.3 & $(0.0, 0.0)$ & 111.24H1.002\\
        2023-06-06 & 2023.4298 & 9 & 0.9 & $(0.0, 0.0)$ &111.24H1.002\\
        2023-07-04 & 2023.5063 & 16 & 1.6 & $(0.0, 0.0)$ &111.24H1.003\\\hline
        2024-03-25 & 2024.2325 & 3 & 0.3 & $(-51.13, 54.59)$& 112.25CV.001\\
        2024-03-27 & 2024.2380 & 2 & 0.2 & $(-51.13, 54.59)$& 112.25CV.001 \\
        2024-03-28 & 2024.2405 & 6 & 0.6 & $(-51.13, 54.59)$ &112.25CV.001\\
        2024-04-29 & 2024.3281 & 4 & 0.4 & $(-56.13, 59.59)$ & 113.268P.001\\
        2024-05-28 & 2024.4073 & 12 & 1.2 & $(-63.08, 63.25)$ & 113.268P.002\\
        2024-06-22 & 2024.4756 & 10 & 1.0 & $(-63.08, 63.25)$ &113.268P.003\\
        2024-06-25 & 2024.4837 & 9 & 0.9 & $(-63.08, 63.25)$& 113.268P.003\\
        2024-07-17 & 2024.5438 & 9 & 0.9 & $(-63.08, 63.25)$& 113.268P.004\\\hline
        2025-03-16 & 2025.2070 & 10 & 1.0 & $(-61., 51.2)$& 114.270V.001 \\
        2025-05-11 & 2025.3602 & 10 & 1.0 & $(-95.26, 51.2 )$& 115.27WT.001	\\
        2025-06-09 & 2025.4393 & 10 & 1.0 & $(-95.26, 51.2)$& 115.27WT.002\\
        2025-07-07 & 2025.5159 & 9 & 0.9 & $(-69.959, 56.581)$&115.27WT.003 \\
        2025-08-06 & 2025.5978 & 7 & 0.7 & $(-72.005, 57.737)$& 115.27WT.004\\
    \end{tabular}
\end{table}

\begin{figure}[ht]
    \centering
    \includegraphics[width=\linewidth]{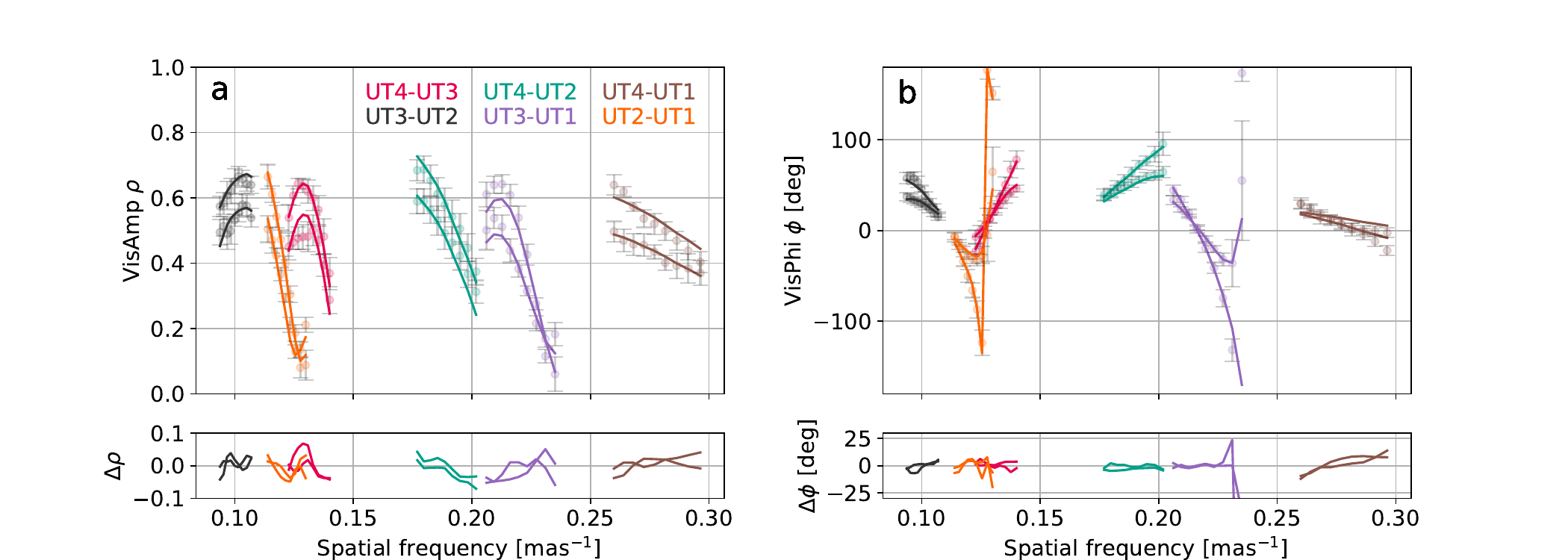}
    \vspace{0.1cm}
    \caption{{\bf Reduced GRAVITY data consisting of complex visibilities.}
GRAVITY GC data, showing the complex visibility samples of a single GRAVITY exposure from 2023-07-04 with $\SI{360}{s}$ integration time, represented as amplitude $\rho$ and absolute phase $\phi$, plotted against spatial frequency. The color scheme encodes the corresponding baseline of the VLTI for two linear polarizations, each with nine spectral channels. The solid lines depict the model as reconstructed from a total of 16 exposures.}
\end{figure}

\begin{figure}[ht]
    \centering
    \includegraphics[width=\linewidth]{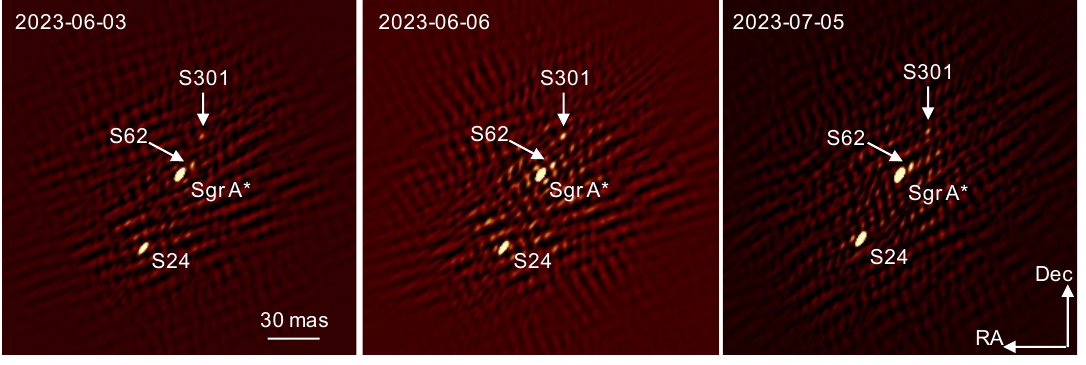}
    \caption{{\bf S301 detections using the CLEAN algorithm.}
Three examples of inferring S301 in 2023 using the CLEAN algorithm implemented in AIPS \protect \citesupp{2003ASSL..285..109G}, following \protect \citesupp{abuter_detection_2021} and \cite{gravity_collaboration_deep_2022}.}
  \end{figure}

\begin{figure}[ht]
    \centering
    \includegraphics[width=0.9\linewidth]{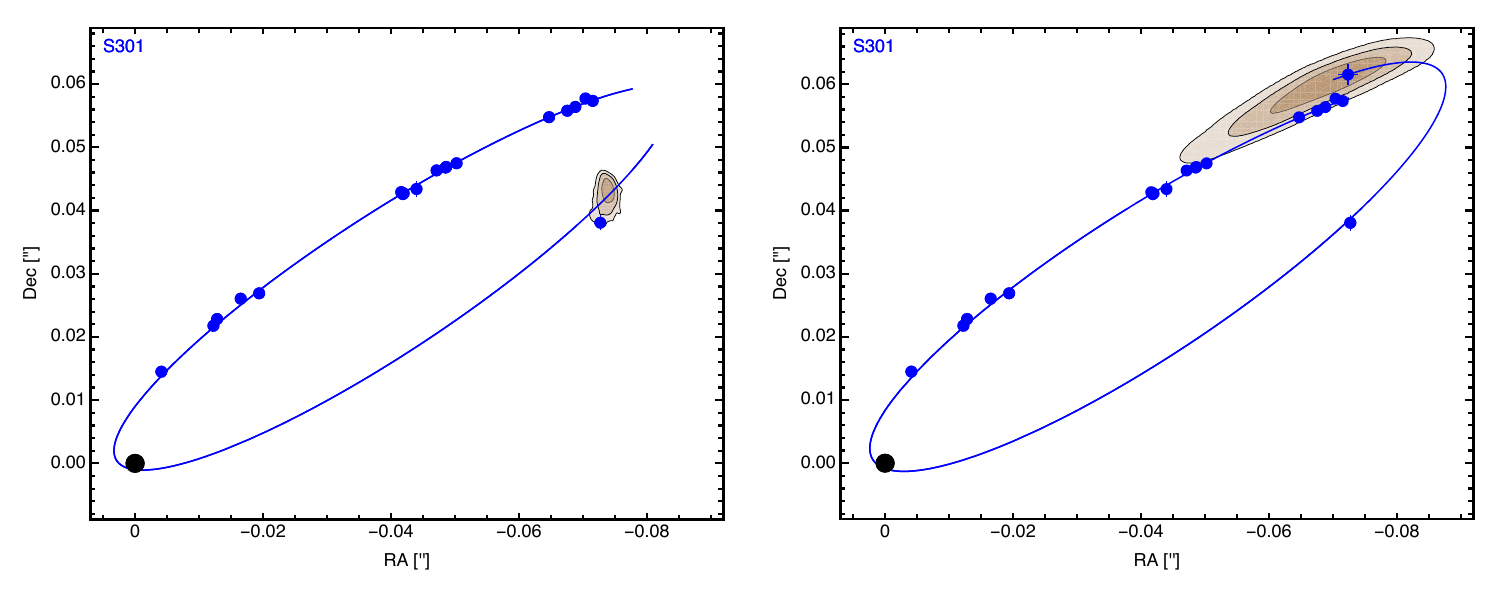}
    \caption{{\bf Post-dicting where S301 is expected in previous epochs.  }
Left: Using the 2023-2025 data, we use samples from a Markov chain to predict the expected position of S301 in 2021 (brown contours). The actual inference falls close to the predicted position. Right: Using now the 2021-2025 data, we repeat the procedure to predict the star's place in 2017. Again, we find S301 where we expect it. }
\end{figure}

\noindent Based on the orbital coverage from 2023-2025, we were able to trace S301 back in time, and identified two previously acquired datasets in 2017 and 2021, in which the star should be present. For this, we fit a preliminary orbit to the 2023-2025 data, and used samples from a Markov chain to predict where the star should be in 2021. The dataset in 2021 was acquired during an observation that involved multiple other pointings in the GC to create a mosaic of the central $\simeq\SI{200}\times\SI{200}{\text{mas}}$ \cite{gravity_collaboration_deep_2022}. This particular pointing was affected by bad seeing conditions, however, and 
we apply stricter cuts and accept as a post-processing step only fringe-tracking ratios of $\SI{90}{\percent}$ for the individual detector integrations, effectively discarding data with low coherent flux. This procedure led to the inference of S301 in this particular epoch, close to the predicted position (Extended Data Fig.~3, left). 

\noindent With that position in hand, we repeated the orbit fit and used the updated Markov chain to predict the S301 position in 2017. The 2017 data were acquired as part of the monitoring of the motion of the star S2, an object around five magnitudes brighter than S301. The dataset consists of individual exposures across five consecutive nights in March 2017, partially recorded in split-polarization mode. In the image reconstruction of this dataset, we treat both polarizations in polarized exposures as individual measurements and combine them with the unpolarized data.
Again,  S301 is found close to the predicted place (Extended Data Fig.~3, right). 

\begin{figure}[ht]
    \centering
    \includegraphics[width=0.5\linewidth]{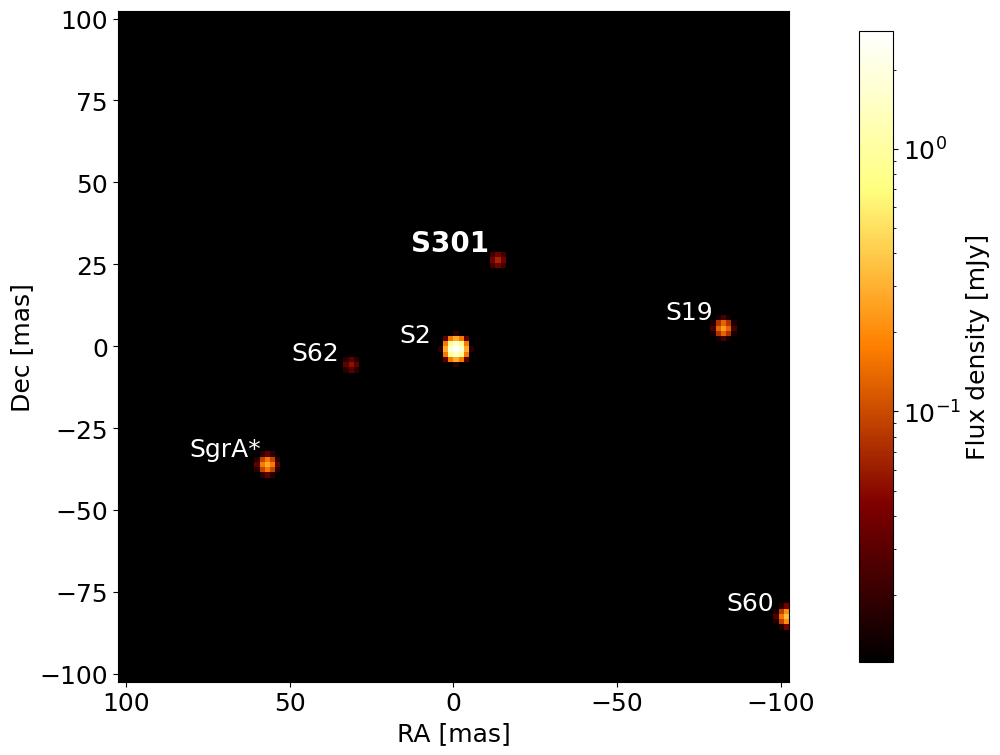}
    \caption{{\bf The 2017 detection of S301.} $\text{G}^\text{R}$ image of the data set in 2017, in which S301 is weakly inferred. }
\end{figure}

\subsection*{Highest-resolution, deep images of the central 800 AU}
\label{subsec_imaging}
The high-angular-resolution, high-fidelity images of the GC are reconstructed from GRAVITY data, an example of which is shown in Extended Data Fig.~1 with the image reconstruction tool GRAVITY-RESOLVE ($\text{G}^\text{R}$) \cite{gravity_collaboration_deep_2022}. This code is designed to reconstruct stars in the GC that appear to GRAVITY as unresolved point sources and, in particular, to find faint, yet undiscovered stars. $\text{G}^\text{R}$ is based on a hierarchical forward model that incorporates the instrument response of GRAVITY including both optical aberrations and the spectral transmission within the beam combiner, as well as a statistical model of the GC.
The intrinsic multitude of degrees of freedom of an image is tamed with Bayesian inference, where the statistical model of the GC provides the prior information. With $\text{G}^\text{R}$, we exploit GRAVITY's supreme imaging resolution of $\simeq\SI{1.7}{mas}$ and its phase-referencing capabilities, which allow for high contrast images and routine mosaicking as pioneered in radio interferometry \cite{mang_in_prep}.

\noindent The individual positions of S301 over time, as depicted in Fig.~2, were inferred in a first step from imaging. We initialized the model in $\text{G}^\text{R}$ by invoking all known, bright sources within the field of view. Positions and fluxes of the stars and Sgr~A* are constrained by prior distributions that reflect current knowledge. We applied $\text{G}^\text{R}$ a total of ten times to the same dataset with varying initial random seeds. Tentative faint sources that may appear in the image of individual reconstructions are only accepted when they are inferred in at least five out of ten reconstructions. Their corresponding positions in the image grid are then referenced to SgrA* and subsequently averaged.

\subsection*{Astrometric errors}
The uncertainties of the newly inferred sources are derived from the scatter over the different runs, taking also into account  the finite resolution of the pixel grid. This is an improvement over  \cite{gravity_collaboration_deep_2022} where the errors were simply approximated by the size of the pixels in the image. If the same faint source is inferred in the same pixel for the ten individual runs, the corresponding standard deviation is zero, albeit unphysical. Hence, the discretized position space needs to be considered when stating errors on astrometry. Instead of quantifying the error based on the number of same-pixel inferences in a set of detections, we opt for a general approach, which may be conservative, but prevents an underestimation of errors. We derive a discretization error by calculating the root mean square error for a single pixel within the image for both directions, RA. and Dec., independently. Considering a pixel size of $\SI{0.8}{mas}$ per pixel, this evaluates to $\approx\SI{207}{\mu as}$ and represents the statistical uncertainty in astrometry originating from the pixel grid in $\text{G}^\text{R}$ images.  This value acts effectively as a noise floor.

\subsection*{Fitting of GRAVITY data}
Besides imaging with $\text{G}^\text{R}$, we apply two other tools to analyze GRAVITY data. These methods serve to fit individual stars, that is, parameterized point sources, foremost to determine their astrometry and photometry. They allow for a crucial cross-check  whether S301 is inferred at the same position with a comparable photometry as with the image reconstruction. Both fitting tools are separately implemented in different programming languages.\\
In a first step, every GRAVITY exposure is fit separately to infer the variable flux of Sgr~A*, equivalent to determining a light curve over time. Both inferred fluxes and positions of Sgr~A* and other bright, known sources then provide the starting values for a subsequent combined fit.  For S301, the values from the imaging code are used as starting values. S301 is indeed inferred by both methods at the same position as with the imaging code, within statistical errors.\\
While the fitting codes yield the same positions for S301 as $\text{G}^\text{R}$, it would be de facto impossible to find a star like S301 using just a fitting code. The fitting essentially returns a local minimum, while the imaging efficiently explores the full parameter space.

\subsection*{Imaging with CLEAN}
We also inferred S301 with a more classical imaging routine, namely CLEAN (Extended Data Fig.~2). For that, we first CLEANed on Sgr~A* in the individual exposures in order to capture its variability. The corresponding coherent flux is then subtracted from the data before combining the data and running a deeper CLEAN, allowing to infer fainter sources, such as S301 and S62.

\subsection*{Best-fit orbit}
\label{sec:bfo}
We fit the astrometric positions of S301 like we did for the S2 data in \cite{2020A&A...636L...5G}, taking into account R{\o}mer delay and the 1PN correction of the motion due to the Schwarzschild nature of the gravitational potential. Relativistic Doppler effect and gravitational redshift do not matter as they only affect radial velocity measurements, which we currently do not have.
 This lack also results in an ambiguity in the three-dimensional orientation of the orbit. The two viable orbit solutions to the proper motion of S301 are given in Extended Data Table~2 (angle conventions follow \cite{gillessen_monitoring_2009}) and leave the semi-major axis, eccentricity, and time of periastron unchanged within errors. In principle, the R{\o}mer delay could break the ambiguity \citep{abuter_geometric_2019}, but does not yet for the limited phase coverage of S301. 
 The orbit stands out compared to other S-stars. For an illustration see Extended Data Fig.~6 and~7.  S301's high eccentricity, combined with the small value of the semi-major axis makes S301 an outlier with the smallest value of $r_p=a (1-e)$. Also note that the orbital plane of one of the two possible solutions agrees to within $3^\circ$ with the inner clockwise disk of young, massive stars in \citesupp{2022ApJ...932L...6V}.

\begin{table}[ht]
\centering
 \caption[]{{\bf Best-fit orbit parameters for S301}. The two sets give the two possible orbit orientations. }
{
 \begin{tabular}{l|cc| cc}
 Parameter&Value&formal fit error & Value&formal fit error\\
 \hline
$a$ [mas] & 83.0 & 0.7 & 83.0 & 0.7\\
$e$  & 0.9832 & 0.0010 & 0.9824 & 0.0011\\
$i$ [$^\circ$] & 124.09 & 1.10 & 122.84 & 1.12 \\
 $\Omega$   [$^\circ$] & 73.8 & 3.5 & 256.9 & 3.3  \\
  $\omega$   [$^\circ$] & 293.4 & 2.2 & 115.1 & 2.0 \\
   $t_P$  [yr] & 2023.126 & 0.010 & 2023.125 & 0.011 \\
 \hline
 P [yr] & 8.68 & 0.11 & 8.68 & 0.11 \\
  \hline
  $\chi^2$& 25.86 && 25.53 & 
 \end{tabular}
 }
\end{table}

\begin{figure}[ht]
    \centering
    \includegraphics[width=0.95\linewidth]{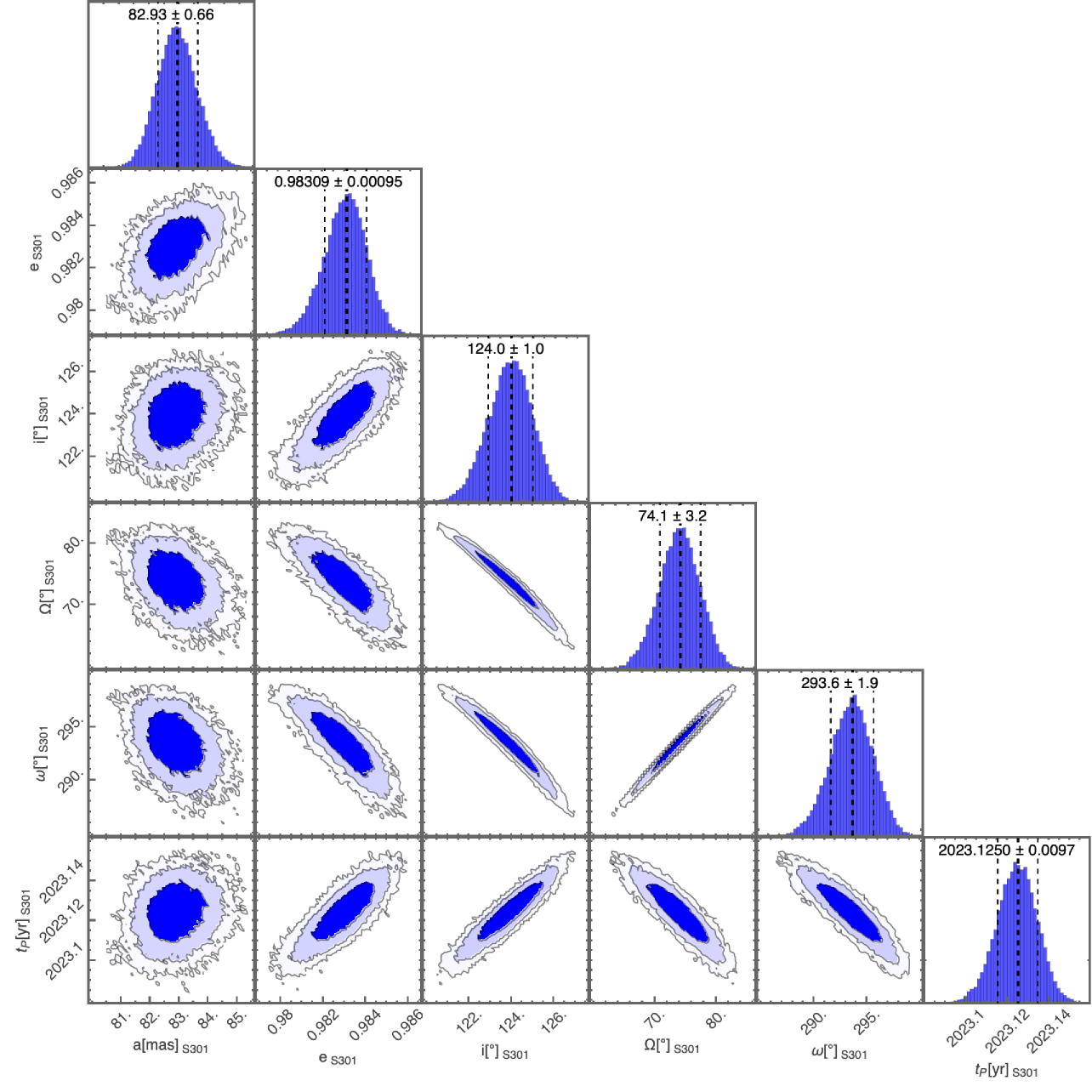}
    \caption{{\bf  Posterior of the best-fit parameters.} Posterior of the six parameter orbit fit in Extended Data Table~2, left column.}
   \end{figure}

\begin{figure}[ht]
    \centering
    \includegraphics[width=0.65\linewidth]{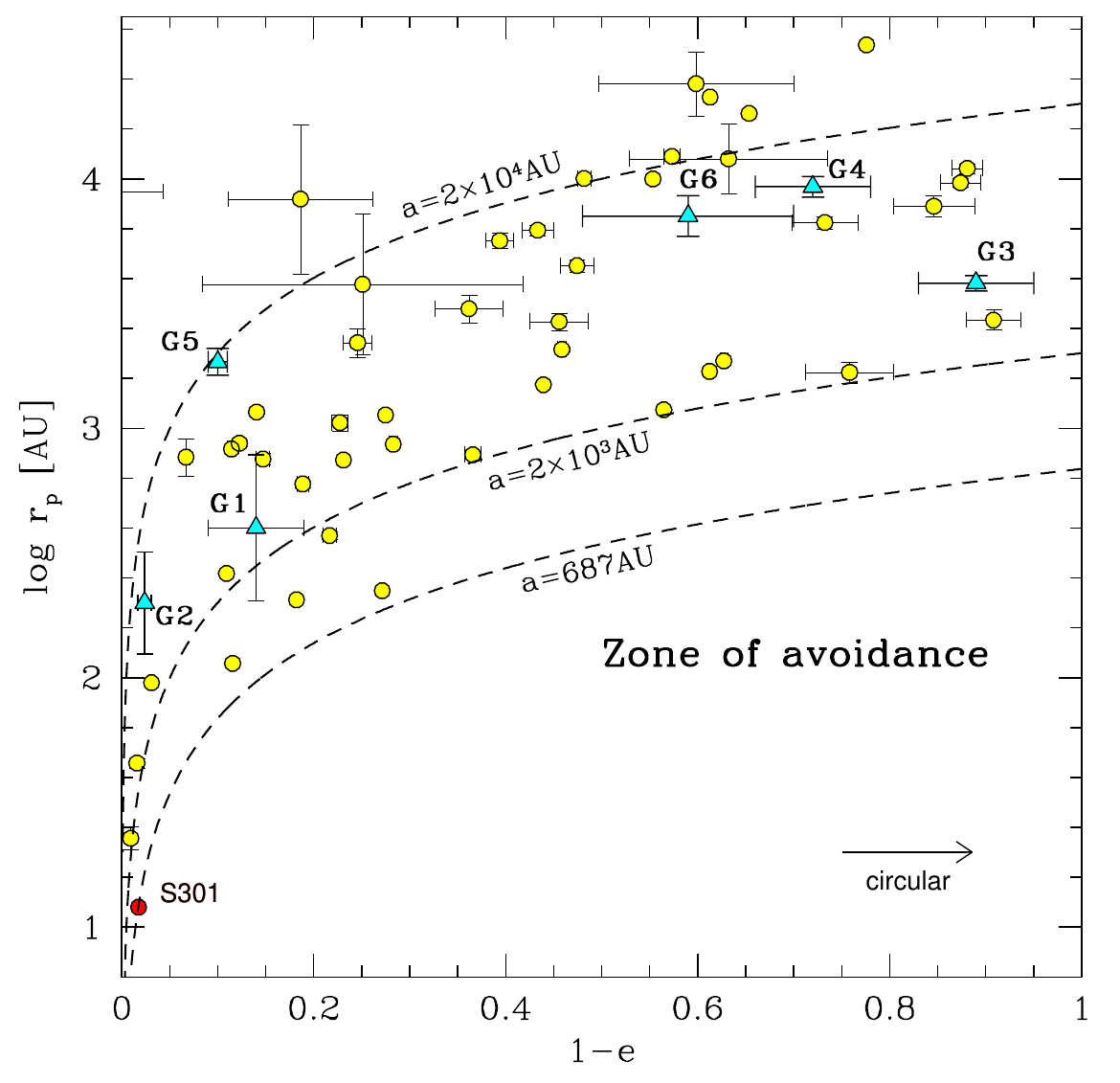}
    \caption{{\bf S301 and the 'zone of avoidance'.} Pericenter distance $r_p$ versus $(1-e)$, with $e$ the eccentricity of S-stars (yellow points) and G-clouds (cyan triangles) orbiting the central MBH \protect \citesupp{2024ApJ...962...81B, generozov_zone_2025}.  The dashed lines connect points of constant $a$.
    The orbits of all stars in the lower right, empty region should have been determined by now. The lack of stars in this so-called 'zone of avoidance' results from the fact that no stars exist in this regime because of a correlation between semi-major axis $a$ and $(1-e)$, such that stars on more circular orbits with smaller values of $e$ have larger semi-major axes. S301 is depicted by the red point in the lower left corner. }
   \end{figure}

\begin{figure}[ht]
    \centering
    \includegraphics[width=0.55\linewidth]{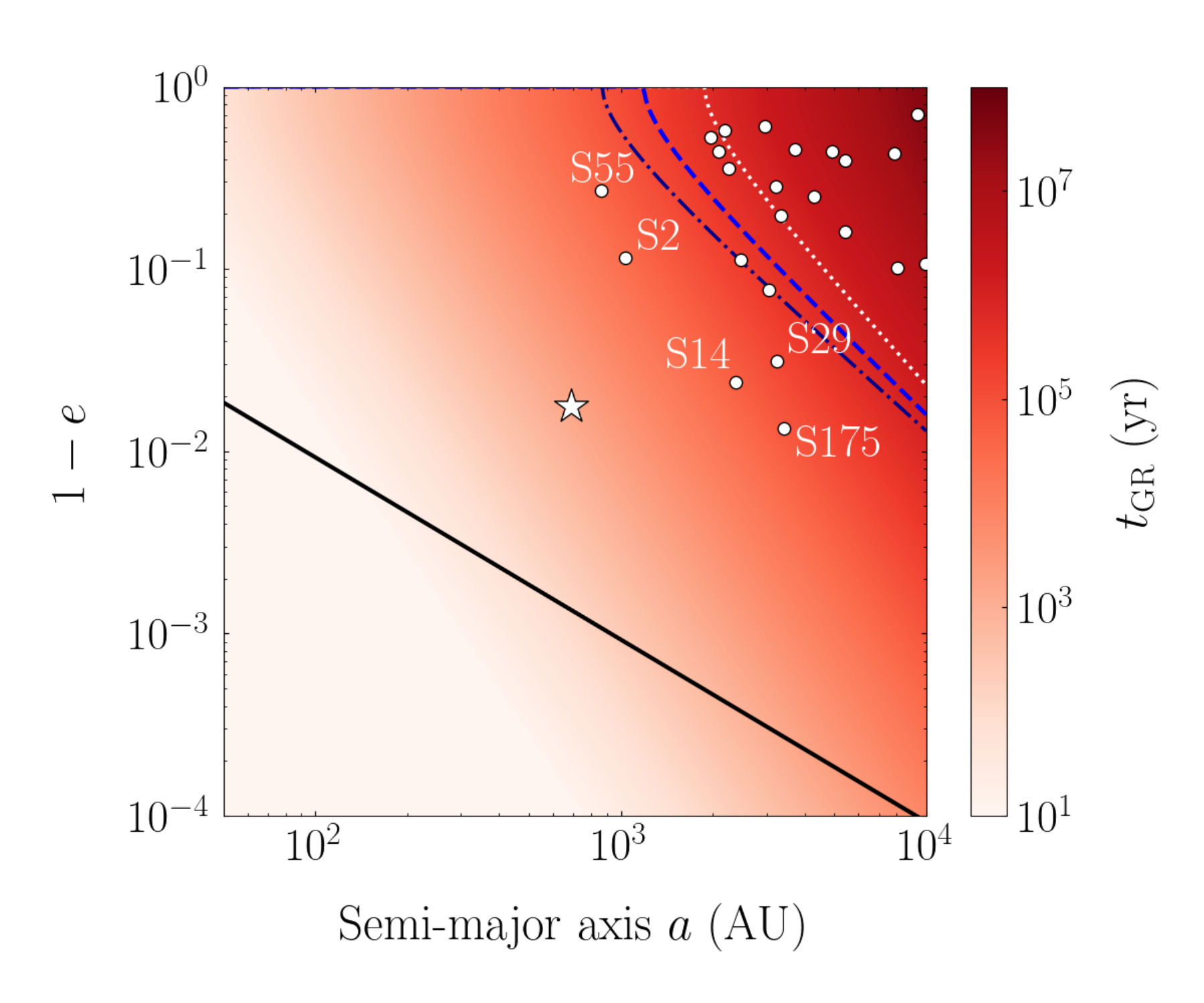}
     \caption{{\bf Dynamical landscape in the $a$ versus $(1-e)$ phase space (in AU)}. The color map indicates $t_{\mathrm{GR}}$. The Schwarzschild Barrier (SB) boundary ($t_{\mathrm{GR}}=t_{\mathrm{RR},v}$) shifts significantly depending on the cusp slope $\gamma$. Boundaries are shown for the observed shallow cusp ($\gamma=1.3$, dotted curve) and steep theoretical cusps ($\gamma=1.75$, dashed curve; $\gamma=2.0$, dash-dotted curve), assuming normalization at 0.25 pc. S301 (white asterisk) and other young S-stars (white circles) remain deep within the barrier in all scenarios. The solid line indicates the tidal disruption limit for S301 ($1.5\,M_\odot, 1.4\,R_\odot$).}
\end{figure}

\subsection*{Measuring the spin of Sgr~A* with S301}
\label{mock}
The orbit-averaged Lense-Thirring 
effects for the in-plane and out-of-plane {major axis} precession are {respectively} \citesupp{2010PhRvD..81f2002M,AbdElDayem2026}
\begin{eqnarray}
\Delta \varpi_\mathrm{LT} &=& - 8 \pi \chi \cos \xi \left( \frac{G M_\mathrm{MBH}}{a(1-e^2) c^2} \right)^{3/2} \,\,, \nonumber \\
\Delta \Theta_\mathrm{LT} &=& -4 \pi \chi \sin \xi \sin \lambda   \left( \frac{G M_\mathrm{MBH}}{a(1-e^2) c^2} \right)^{3/2} \,\,,
\end{eqnarray}
where $\xi$ is the inclination between spin axis and orbital angular momentum, and $\lambda$ is the position angle of the projection of the spin axis onto the orbital plane. For S301, the in-plane contribution per revolution
amounts to $0.11^\circ \chi \cos \xi$.

\noindent In order to assess how S301 probes the spin of Sgr A*, we perform a mock data analysis combining current and simulated future observations. We construct synthetic astrometric and radial-velocity datasets extending the existing measurements with simulated observations between 2026 and 2035, using the values of the orbital parameters given in Extended Data Table~2, with mass and distance of Sgr A* from \cite{gravitycollaborationMassDistributionGalactic2022}. We optimistically adopt $\chi = 1$ and an orientation approximately aligned with the stellar orbital angular momentum. In this case, the Lense-Thirring effect produces mainly in-plane precession and little precession of the orbital plane. The resulting mock data is fitted to test whether the spin parameters can be recovered. We assume a realistic astrometric precision of $\SI{100}{\mu as}$ (as expected for the final performance of GRAVITY+) and $1\,$km/s accuracy on the radial velocity as reachable with future ELT/MICADO observations. A sampling of 10 data points per year until 2035, with 10 additional data points around pericenter, would yield a spin constraint with an uncertainty on $\chi$ of $<0.2$ (Fig.~3, right), equivalent to a $>5 \sigma$ detection of $\chi = 1$ compared to the non-spinning ($\chi = 0$) case. In Extended Data Fig.~9, we show the dependence of the significance of the spin detection in our simulations as a function of orientation of the black hole spin.
The maxima of significance correspond to (anti-)alignment of the spin and orbital angular momentum, while the minima correspond to alignment of the spin along the semi-major axis of the orbit. A more detailed discussion is provided in \citesupp{AbdElDayem_inprep}.  Also, we note that future robust spin constraints will require modeling up to 2PN order to avoid systematic biases at low spins.

\subsection*{Spectral type, mass and age of S301}
\label{app:age}
Given the magnitude  $m_K=19.3 \pm 0.3$ and assuming an extinction of 2.42 \citesupp{2011ApJ...737...73F} and a GC
distance of $R_0=8.3\,$kpc \citep{abuter_geometric_2019}, S301 has an absolute K magnitude of  $\approx 2.28$.  It is too faint to be a giant, but it instead is compatible with being a main-sequence star.  Its 
spectral type then is a late A-type or an early F-type star (see also Fig.~2 in \citesupp{2010RvMP...82.3121G}), which means its color index is $V-K = 0.6$ \citepsupp{allen2000}. With an  absolute V magnitude of 2.88 it has $L=5.5\,L_\odot$ and a spectral type of F1.5, corresponding to a mass of just below $1.5\,M_\odot$.
Alternatively,  its mass can be estimated to be between $\sim1.1 M_{\odot}$ and $\sim1.5 M_{\odot}$, depending on the age of the star, with younger ages corresponding to larger masses. 

\noindent We infer the possible ages and masses of this star, using the observed brightness and 
the MIST \citesupp{dotter2016, choi+2016, paxton+2015} isochrones\footnote{MIST Version 1.2 tracks with $\Omega/\Omega_{\rm crit}=0.4$ and solar metallicity}.
Specifically, we identify evolutionary points where the track magnitude crosses the inferred absolute magnitude. 
The stellar ages and masses for such points are shown as a solid, blue line in Extended Data Fig.~8. Note that the MIST tracks do not include the K band magnitude, and we use the JWST F210M magnitude as a proxy. 

\noindent To test the robustness of our results we repeat this analysis with the K band magnitude from
 \textsc{PARSEC} isochrones (v1.2S, \citesupp{bressan+2012, chen+2015, tang+2014, marigo+2017, pastorelli+2020}), and show the possible ages and masses as a dashed, orange line in Extended Data Fig.~8. We note that unlike MIST, the \textsc{PARSEC} tracks do not include stellar rotation. 
Nonetheless, the results are in excellent agreement, with the stellar mass ranging from $\sim1.1 M_{\odot}$ to $\sim1.5 M_{\odot}$, depending on the age of the star.

\begin{figure}[ht]
    \centering
    \includegraphics[width=0.45\linewidth]{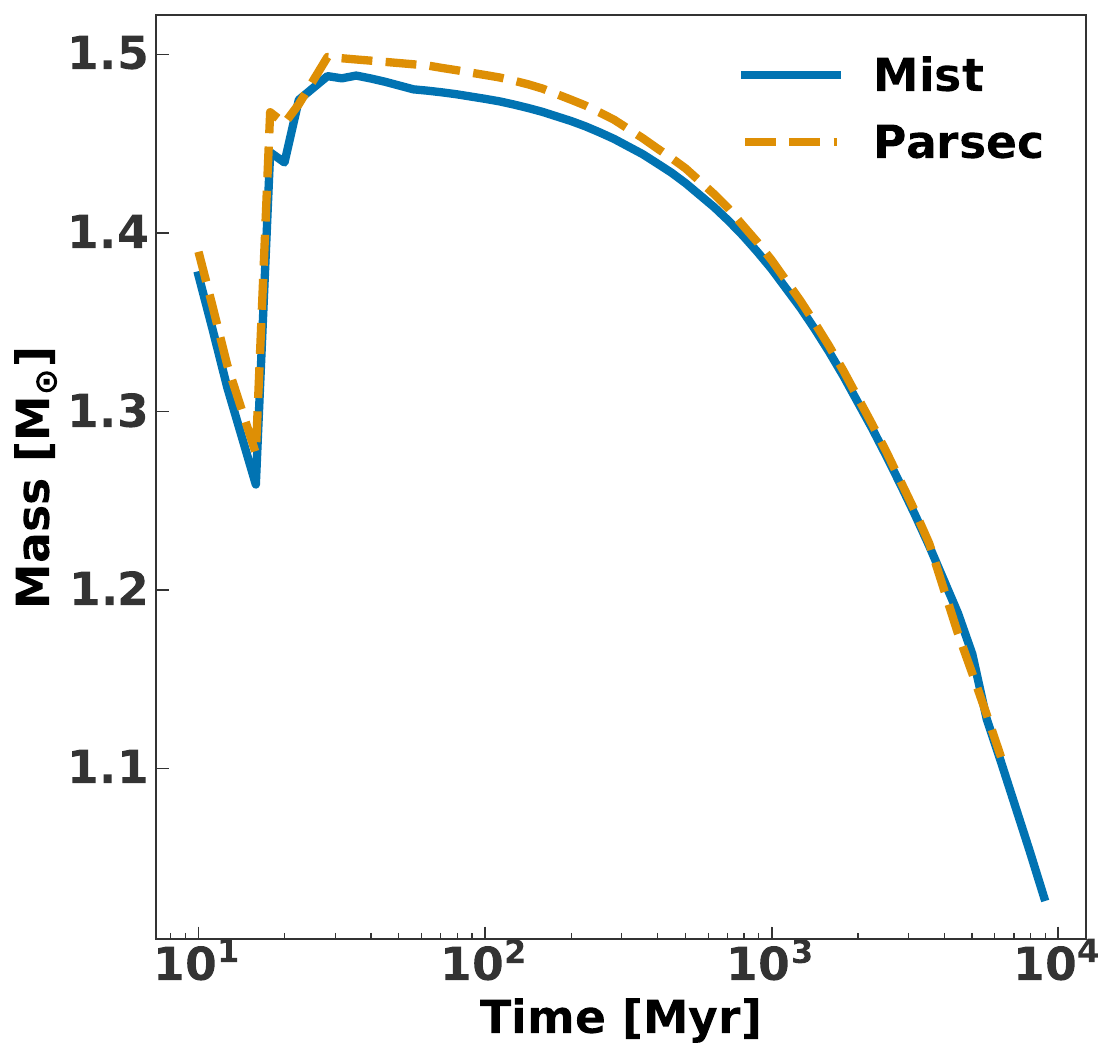}
     \caption{{\bf Possible ages and masses for S301.} Possible ages and masses for S301, from its observed K-band magnitude and theoretical isochrones. }
\end{figure}

\begin{figure}[ht]
    \centering
    \includegraphics[width=0.9\linewidth]{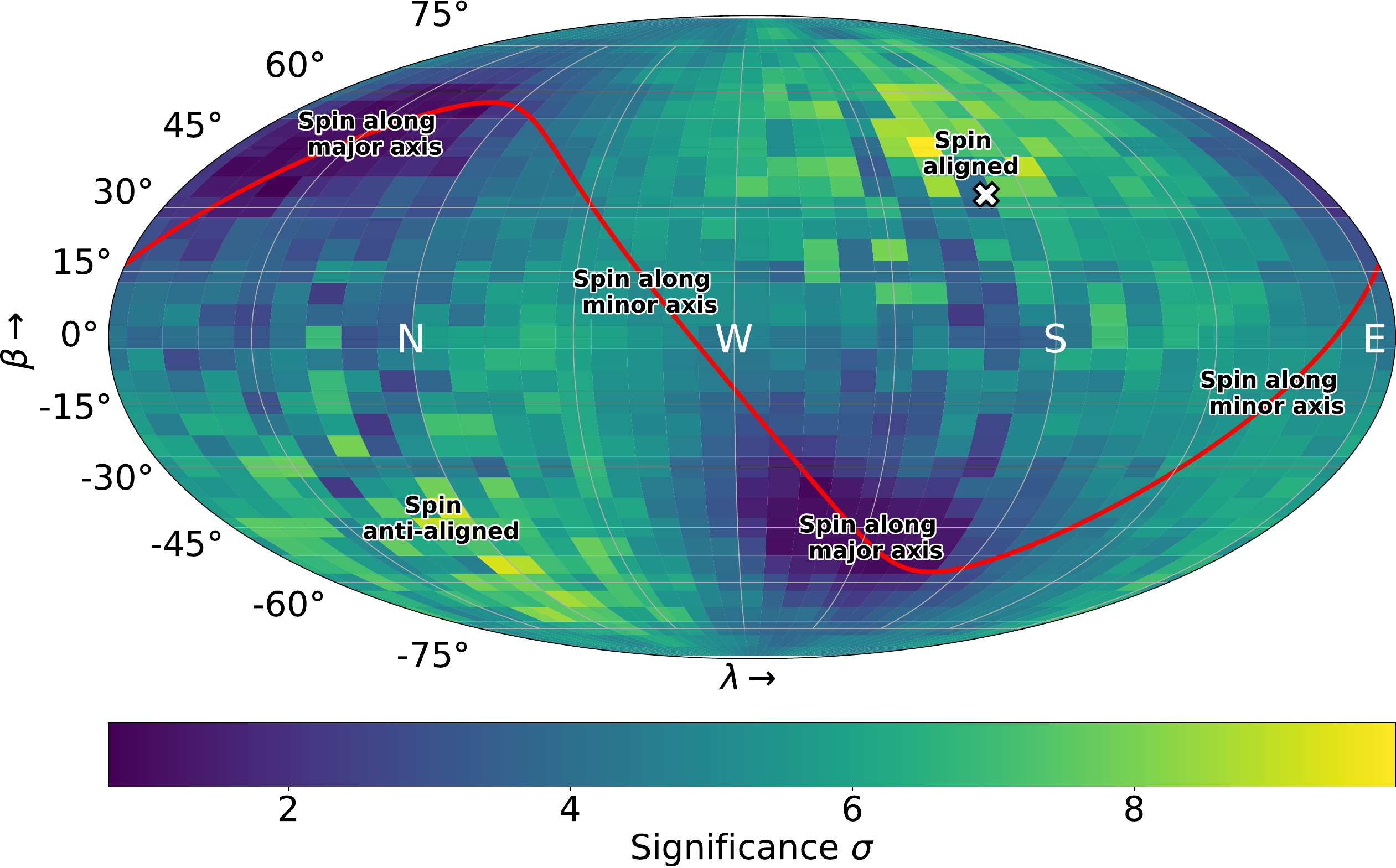}
     \caption{{\bf Sensitivity of the S301 orbit to the spin orientation.}
Significance of the detection of the spin as a function of orientation of the spin vector for our simulated data with $\chi=1$. Over a wide range of possible orientations, we expect to be able to detect the spin. The cross marks the orientation of the orbital angular momentum vector of S301. The red line corresponds to maximum misalignment between black hole spin and orbit. The best chances of detection are for co- or anti-alignment.}
\end{figure}

\subsection*{Tidal effects}
\label{app:tidal}
Interestingly \citesupp{2013ApJ...777...57P} suggest that  tidal effects will prevent observations of Kerr effects around Sgr~A*. However, they consider stars 10 $M_\odot$ or heavier. S301 is sufficiently small so that these tidal effects are negligible. The energy input per orbit can be estimated, using the \citesupp{press&teukolsky1977} formalism, viz.
\begin{align}
     &\Delta E \approx T_2(\eta) \left(\frac{M_{\mathrm MBH}}{m_\star}\right)^2 \frac{G m_\star^2}{R_\star} \left(\frac{r_p}{R_\star}\right)^{-6} \nonumber \\
     &\eta = \left(\frac{m_\star}{M_{\mathrm {MBH}}}\right)^{1/2} \left(\frac{r_p}{R_\star} \right)^{3/2},
 \end{align}
where $r_p$ is the pericenter distance, $R_\star$ the stellar radius, $m_\star$  the stellar mass, and $T_2$  the tidal coupling constant. Here, we only include the leading order quadrupole term. We estimate the tidal coupling constant using the fits from \citesupp{1993A&A...280..174P}, assuming an $n=3$ polytrope. 
Note that these fits only extend to $\eta=10$, and we extrapolate them using the logarithmic slope there. Thus,
\begin{equation}
\label{e3}
    T(\eta)\approx 3.8 \times 10^{-5} \left(\frac{\eta}{10}\right)^{-6.5}
\end{equation}
for $\eta \geq 10$.  This yields $\delta E \approx 10^{-16} G m_\star^2 / R_\star$. It would take of order $10^{17}$ yr for tides to inject an order unity fraction of the energy of the star. Therefore, tidal heating can be neglected at the star's current orbit.

\subsection*{Comparison with previously claimed short-period stars}
\label{app:comparison}
Over the past five years, a number of short-period stars with orbital periods as low as 4 years were claimed to have been discovered by one team using adaptive-optics based imaging techniques \citesupp{2020ApJ...899...50P,  2022ApJ...933...49P}, thus at a resolution 15 times worse than the GRAVITY data.  Safely, we can exclude that the objects in \citesupp{2020ApJ...899...50P} named S4711 and S62 (different from the star our team calls S62, see  \citesupp{abuter_detection_2021}), and that have similar orbital periods as S301, are actually S301:
\begin{itemize}
    \item The eccentricity of S4711 with $e=0.768$ is considerably less than S301's, and the claimed orbit has a different projection on sky, extending towards the East.
    \item While S62 features a comparable eccentricity of $e=0.976$, the claimed orbit revolves counter-clockwise, unlike S301.
\end{itemize}
Beyond that, none of the other claimed stars have orbital elements similar to the ones of S301. Hence, S301 is newly discovered. 

\noindent Further it is worth noting that the limiting magnitude of $m_K \sim 20$ inferred from the GRAVITY observations and reported here should have allowed us to easily detect any of the claimed discoveries if covered in our exposures. 
In none of our reconstructed images we have detected a significant contribution of flux beyond background noise that one could attribute to the claimed stars.

\subsection*{Newtonian perturbations of the spin measurement}
\label{perturbations}
The simulations for the spin measurement assume that S301 orbits an isolated Kerr black hole in the absence of external perturbations. As noted by \citesupp{2010PhRvD..81f2002M}, gravitational perturbations from a population of stellar-mass black holes, expected to form a power-law-density stellar cusp around Sgr~A*, can induce orbital precession comparable in magnitude to the Lense-Thirring effect, thereby complicating the measurement of the spin. Recent constraints from \cite{gravity_collaboration_mass_2024}  limit the extended mass within the central 10 mpc to $\lesssim 1200\,M_\odot$. To assess the impact of a realistic perturber population, following \citesupp{2025A&A...701A..89S} we performed N-body simulations of S301 including a cluster of 60 stellar-mass black holes of $20\,M_\odot$ each, distributed within 10 mpc of Sgr~A* according to a density profile $\rho(r)\propto r^{-2}$. Considering 100 independent realizations of the initial conditions, we find that the cluster induces an average orbital-plane precession per orbital period of $0.65^{+0.36}_{-0.56}$ arcminutes. This contribution is typically subdominant compared to the Lense-Thirring precession for moderate-to-high dimensionless spins and favorable orientations of the black-hole spin relative to the orbital angular momentum. Importantly, the Lense-Thirring effect is concentrated near pericenter, whereas perturbations from a granular stellar background tend to produce their largest observable deviations near apocenter \citesupp{2022A&A...660A..13H,2025A&A...701A..89S}. This phase separation, together with the distinct temporal signatures of the two effects, offers a promising avenue to disentangle the relativistic spin signal from stellar perturbations and thereby enable a robust measurement of the MBH spin.

\noindent In an extreme case in which all the mass (allowed by S2 observations) is concentrated within the orbit of S301 in a disk and in which the orientation of the orbit is a few degrees from the disk, the nodal precession due to the disk might be larger, by up to an order of magnitude, than the Lense-Thirring precession. However, for most orientations and for less extreme mass distributions, the Newtonian effect of this mass on the nodal precession is expected to be one or two orders of magnitude smaller than the Lense-Thirring effect.

\subsection*{A Hills origin for S301}
\label{app:hills}
For the Hills mechanism \citep{hillsHypervelocityTidalStars1988}, \citesupp{yu_tremaine_2003,gou+03,Gin+06,PeretsHopmanAlexander2007,2020ApJ...896..137G} one of the binary components is captured at a semi-major axis of 
\begin{equation} 
a_{\rm cap} = f_1 \left(\frac{M_\mathrm{MBH}}{m_{\rm bin}}\right)^{2/3} a_{\rm bin},
\label{e2}
\end{equation}
where $f_1 \approx 0.5$ for circular, equal mass binaries \citesupp[e.g.][]{hills1991}.
The captured star will inherit the pericenter of the binary's orbit around the MBH, and so the binary separation must be no more than a factor few times  $\left(M_\mathrm{MBH}/m_{\rm bin}\right)^{1/3} a_{\rm bin}$ (the characteristic binary tidal disruption radius). Thus, the eccentricity of the captured star will be
\begin{equation}
    e_\mathrm{cap} = 1 - f_2 \left( \frac{m_{\rm bin}}{M_\mathrm{MBH}} \right)^{1/3},
\end{equation}
where $f_2$ is at most a factor of order unity. The expected distribution of $f_2$ and hence of $e_\mathrm{cap}$ will depend on the distribution of binary properties (e.g. the mass ratio, inclination, internal eccentricity) and the pericenter distribution of disrupting binaries. Recently, \citesupp{generozov_zone_2025} simulated binary disruptions in the GC for an observationally motivated binary population, assuming a full loss cone and an isotropic inclination distribution. For captured stars with semi-major axes between $2\times 10^{-3}\,$pc and $4\times 10^{-3}\,$pc and masses between $1.3 M_{\odot}$ and $1.7 M_{\odot}$ the eccentricities ranged between 0.97-0.996 (5th-95th percentile) with a median eccentricity of 0.985.
Thus, binary disruption would naturally explain the observed eccentricity of the star. 

\noindent The measured semi-major axis constrains the pre-disruption binary. For nearly equal masses the Hills mapping gives $a_\mathrm{cap}\sim \frac{1}{2}(M_\mathrm{MBH}/2m_\ast)^{2/3}a_\mathrm{bin}$ \citesupp[e.g.][]{hills1991}. Identifying $a_\mathrm{cap}\simeq a$ and adopting $m_\ast\simeq 1.3$--$1.7\,M_\odot$ yields a pre-disruption separation $a_\mathrm{bin}\sim 0.05$--$0.2~{\rm AU}$, corresponding to orbital periods $P_\mathrm{bin}\sim 5$--$20$~days. We find a similar range of progenitor semi-major axes in the simulations of \citesupp{generozov_zone_2025}, with an observationally motivated mass ratio distribution.

\noindent Such compact binaries are common among F-type stars \citep{raghavan_binary_2010} and are expected to be tidally circularised and nearly synchronised. If S301 was initially synchronised, its spin period at capture would have been comparable to $P_\mathrm{bin}$, implying an equatorial rotation velocity $v_\mathrm{rot}\sim 20$--$70~{\rm km\,s^{-1}}$ for $R_\ast\simeq 1.3$--$1.7\,R_\odot$. 
High-resolution spectroscopy with  ELT/MICADO might therefore provide a direct, testable prediction of the Hills-binary origin via a measurement of $v\sin i$.

\noindent Hills events simultaneously populate the innermost S-star cluster and the halo hyper-velocity star population.
 For Hills disruption rates of order $\dot{N}_\mathrm{Hills}\sim 10^{-5} - 10^{-4}\,\mathrm{yr^{-1}}$, simple steady-state arguments suggest that one expects of order a few S301-like stars on similarly relativistic orbits at any given time. 
The Hills disruption rate is supported by both the observed number of hyper-velocity stars in our galaxy \citesupp{2015ARA&A..53...15B} and theoretical estimates as well as the observed tidal disruption rate in other galaxies. With roughly 10\% of them having the right separation range to produce S301 like orbits (see equation~\eqref{e2} above), we take the formation rate of stars on S301 like orbits to be $\dot N_{S301} \sim 10^{-6}\,\mathrm{yr^{-1}}$. Given the collisional lifetime of more than $10^{8}$ years (see Extended Data Table~3), we expect in steady state about a hundred stars with such an orbit. Most of them are likely solar mass stars, and hence still too faint to be detected. Further analysis of the observed coverage and current sensitivities of GRAVITY+ is needed to assess if this estimate is consistent with only one S301-like object detected so far; and continued GRAVITY+ monitoring and future ELT observations may therefore reveal a population of faint, low-mass S-stars deep in the potential well, enabling ensemble constraints on the spin of Sgr~A* and on the distribution of stellar and remnant perturbers in the innermost $\sim 10^{-2}\,\mathrm{pc}$.

\subsection*{Dynamical time scales for S301}
\label{app:Timescales}
In this section we will estimate the time scales for the orbital relaxation and collisions for S301. 

\subsubsection*{Relaxation times}
Relaxation is the dynamical evolution of objects due to perturbations from their environment. In the GC relaxation can be subdivided into resonant and non-resonant parts. Resonant relaxation corresponds to the evolution of stellar orbits due to coherent torques from the background. This can lead to rapid evolution of angular momentum. On longer time scales, non-resonant (two-body) relaxation evolves the orbital energy due to uncorrelated two-body encounters. (See \citesupp{alexander_2017_review} for a review.)

\noindent The two-body relaxation time scale is approximately (see equation 5.61 in \citesupp{merritt_textbook_2013})
\begin{align}
    &t_{rx} = 0.34 \frac{\sigma^3}{G^2 n \langle m^2\rangle \ln \Lambda} 
    \label{eq:trx}
\end{align}
where $\sigma \approx \sqrt{G M/ (1 + \gamma)/r}$ is the (1D) velocity dispersion, $n$ is the number density (with power-law index $-\gamma$). $\langle m^2 \rangle$ is the second moment of the mass function. Due to the quadratic mass dependence, the most massive species (viz. stellar mass black holes) often dictate this time scale. 

\noindent We follow \citesupp{vasiliev2017} to estimate the background density profile. We model the background as two species: 10 $M_{\odot}$ black holes and 1 $M_{\odot}$ stars. This is a reasonable approximation for modeling relaxation in an evolved galactic nucleus,
and has been used extensively in the literature \citesupp{merritt_textbook_2013}. The stars are initialized with a Nuker density profile \citesupp{lauer_nuker_1995}, viz.
\begin{equation}
    \rho_o \left(\frac{r}{r_o}\right)^{-\gamma} \left[1 + \left(\frac{r}{r_o}\right)^\alpha\right]^{(\gamma - \beta) / \alpha},
\end{equation}
~\\
\noindent with $\gamma$=1.5, $\alpha$=2, and $\beta$=5. The density normalization is set by the total mass: {${2.5\times\,10^7\,M_{\odot}}$} and $2.5 \times 10^5 M_{\odot}$ for the stars and black holes respectively. The central MBH starts at $4.15\times 10^6 M_{\odot}$ and grows to $4.27\times 10^6 M_{\odot}$ by consuming stars and black holes. The profile is allowed to relax for 10 Gyr, such that the final density profiles are not too sensitive to the initial conditions. 
In the end the total mass within 0.01 pc is approximately $1200 M_{\odot}$, consistent with the latest constraints on the enclosed mass within S2's apocentre \citep{gravity_collaboration_mass_2024}. Furthermore the final stellar density at 1 pc ($\approx 9\times 10^4 M_{\odot}$ pc$^{-3}$) is comparable to observational estimates \citesupp{schodel_diffuse_light_2018}.

\noindent The following power-law fits approximate the final density profile between $\sim 10^{-4}$ and 0.01 pc
\begin{align}
&\rho_{\rm bh} = 4.95 \times 10^8  \left(\frac{r}{0.0033 \,{\rm pc}}\right)^{-1.75} M_{\odot}\, {\rm pc}^{-3} \nonumber\\
&\rho_{\rm *} =  3.46 \times 10^8 \left(\frac{r}{0.0033 \,{\rm pc}}\right)^{-1.4} M_{\odot}\, {\rm pc}^{-3},
\label{eq:background}
\end{align}
Combining equations~\eqref{eq:trx} and~\eqref{eq:background} we find that the two-body relaxation time scale at S301's semi-major axis ($3.3\times 10^{-3}$ pc) would be $\sim 9\times 10^8$ yr. 

\noindent To estimate the angular momentum relaxation time, including resonant relaxation, we follow \citesupp{tep+2021} (and the references therein). In particular we use their software, JuDOKA\footnote{\url{https://github.com/KerwannTEP/JuDOKA}}, to compute diffusion coefficients ($D_{\rm jj}$) as a function of eccentricity and semi-major axis for the density profiles in equation~\eqref{eq:background}. 

\noindent The angular momentum relaxation times can then be estimated using 
\begin{align}
    &t_{\rm rx, j (NRR)} = \frac{j^2}{D_{\rm jj, NRR}}\nonumber \\
    &t_{\rm rx, j (RR)} = \frac{j^2}{D_{\rm jj, NRR} + D_{\rm jj, RR}},
\end{align}
where $j$ is the angular momentum normalized to the circular angular momentum at the same energy. The top (bottom) row corresponds to the time scale without (with) resonant relaxation. We find that both $t_{\rm rx, j (NRR)}$ and $t_{\rm rx, j (RR)}$ are $\approx 3 \times 10^7$ yr for S301, indicating that resonant relaxation is unimportant for this star. This is expected, because of the star's short Schwarzschild-precession time scale. In other words, the star lies within the Schwarzschild barrier, where rapid precession suppresses the build-up of coherent torques \citesupp[e.g.][]{merritt_schwarzschild_barrier_2011}.   

\noindent Alternatively, $t_{\rm rx, j (NRR)} \approx j^2 t_{\rm rx}$, which gives results that are consistent with equation~\eqref{eq:trx}.

\subsubsection*{Collision time scales}
The mean time between collisions for a single target star is
\begin{equation}
    t_{\rm coll} = \frac{1}{n\,\Sigma\,v_{\rm rel}} ,
    \label{eq:t_coll_def}
\end{equation}
where $n$ is the local number density of potential impactors, $v_{\rm rel}$ is the typical relative velocity, and $\Sigma$ is the collisional cross section. For a star of radius $R_\star$ and mass $m_\star$ colliding with objects of mass $m_{\rm imp}$ and radius $R_{\rm imp}$, the cross section including gravitational focusing is
\begin{equation}
    \Sigma = \pi \left(R_\star + R_{\rm imp}\right)^2 \left(1 + \frac{v_{\rm esc}^2}{v_{\rm rel}^2}\right),
    \qquad
    v_{\rm esc}^2 = \frac{2G\,(m_\star + m_{\rm imp})}{(R_\star + R_{\rm imp})}.
    \label{eq:sigma_general}
\end{equation}
In the case of S301, $m_\star\approx 1.5 M_{\odot}$ and $R_\star \approx 1.4 R_{\odot}$. 
Then from equations~\eqref{eq:background} and~\eqref{eq:sigma_general} the local collision time scale would be $\sim 2.0\times 10^9$ yr for stellar mass black holes and $\sim 2.1 \times 10^8$ yr for stars.

\subsubsection*{Uncertainties, caveats and other effects}
The relaxation and collision time scales we estimate above are local. In principle, the interactions at pericenter can significantly shorten the energy relaxation and collision time scales. A na\"ive orbit average with the background profile in equation~\eqref{eq:background} would suggest a reduction at the order of magnitude level. In practice, fewer than one scatterer is expected at S301's pericenter for our assumed density profiles, so the local time scales are more realistic. 

\noindent Our time scale estimates implicitly assume stars are evolving through a localized diffusion process. In reality, angular momentum and energy perturbations from a handful of massive perturbers exhibit a heavy power-law tail, such that orbital evolution would be dominated by large, non-local jumps that can speed up the orbital evolution dramatically \citesupp{amaro-seoane_frac_dynamics_2025}. 

\subsubsection*{Summary of timescales}
There is a clear timescale hierarchy at S301's radius, with  $T_{\rm orb} \ll T_{\rm SP} \ll T_{\rm LT} \ll T_{\rm Vec-RLX} \ll T_{\rm coll}, T_{\rm Sca-RLX}  \ll T_{\rm GW}$ (Extended Data Table~3), which allows identifying the processes that are relevant for S301's orbit evolution.

\begin{table*}[ht]
 \caption[]{{\bf Dynamical time scales  for S301 and S2.} $T_\mathrm{orb}$ is the orbital period. $T_\mathrm{SP}$ is the Schwarzschild precession time scale, $T_\mathrm{LT}$ for Lense-Thirring precession \protect \citesupp{KocsisTremaine2011}. $T_\mathrm{Vec-RLX}$ refers to the angular momentum relaxation time scale, estimated both with (`RR') and without (`No RR') resonant relaxation. $T_\mathrm{Sca-RLX}$ is the energy relaxation time scale. $T_{\rm coll, bh}$ and $T_{\rm coll, \star}$ are the collision time scales with background black holes and stars respectively. The collision and relaxation times are estimated for relaxed stellar density profiles. The gravitational wave inspiral time scale $T_\mathrm{GW}$ is given by \protect \citesupp{Peters1964}, and the energy dissipation time scale $T_\mathrm{heat}$ by \protect \citesupp{PressTeukolsky1977} (equation~\eqref{e3}).
 We assume $1.5 M_{\odot}$ and $1.4 R_{\odot}$ for the stellar mass and radius of S301, and $15 M_{\odot}$ and $4.7 R_{\odot}$ for S2. Our estimates of the relaxation times are sensitive to the assumed background density profiles. The relaxation times may be significantly longer if there is a dearth of stellar mass black holes at the radial regime of S301.
 }
 \begin{center}
 {
 \begin{tabular}{l|c| c}
 Time scale [yr]&S301 & S2\\
 \hline
$T_\mathrm{orb}$ & 8.7 & 16.1 \\
$T_\mathrm{SP}$   & $1.6 \times 10^3$ & $2.8 \times 10^4$ \\
$T_\mathrm{LT}$   & $5.8 \times 10^4$ & $3.1 \times 10^6$  \\
$T_\mathrm{Vec-RLX}$ (No RR)   & $2.5 \times 10^7$ & $2.3 \times 10^8$  \\
$T_\mathrm{Vec-RLX}$ (RR)   & $2.0 \times 10^7$ & $3.6\times 10^7$  \\
$T_\mathrm{Sca-RLX}$   & $8.6 \times 10^8$ & $1.1 \times 10^9$  \\
$T_\mathrm{coll, bh}$   & $1.8 \times 10^9$ & $4.0 \times 10^8$  \\
$T_\mathrm{coll, \star}$   & $1.7 \times 10^8$ & $4.6 \times 10^{7}$  \\
$T_\mathrm{GW}$   & $2.9\times 10^{10}$ & $7.1 \times 10^{12}$ \\
$T_\mathrm{heat}$ & $3 \times 10^{16}$   & $2 \times 10^{29}$ 
 \end{tabular}
 }
\end{center}
\end{table*}

\newpage

\bibliographystylesupp{sn-mathphys-num}
\bibliographysupp{sn-bibliography}

\newpage 

\small{
\noindent {\bf Data availability statement.}  
This work is based on observations collected at the European Southern Observatory (ESO) under the ESO programme IDs 60.A-9102(A), 105.20B2.004, 111.24H1.00[123], 112.25CV.001, 113.268P.00[1234], 114.270V.001, 115.27WT.00[1234]. These data are publicly available through the ESO archive (\href{https://archive.eso.org/cms.html}{https://archive.eso.org/cms.html}). The astrometric data are available upon request to the corresponding author S.~Gillessen (ste@mpe.mpg.de).
~\\

\noindent {\bf Code availability statement.}   
The software to reduce the GRAVITY data is publicly available (\href{https://www.eso.org/sci/software/pipelines/gravity/}{https://www.eso.org/sci/software/pipelines/gravity/}). The $\text{G}^\text{R}$ source code is available at \href{https://gitlab.mpcdf.mpg.de/gravity/gr_public.git}{https://gitlab.mpcdf.mpg.de/gravity/gr\_public.git}. 
~\\

\noindent {\bf Acknowledgements.}  
We are very grateful to our funding agencies (MPG, ERC, CNRS [PNCG, PNGRAM], DFG, BMBF/BMFTR, Paris Observatory [CS, PhyFOG], Observatoire des Sciences de l'Univers de Grenoble, and the Funda\c c\~ao para a Ci\^encia e a Tecnologia), to ESO and the Paranal staff, and to the many scientific and technical staff members in our institutions, who helped to make NACO, SINFONI, ERIS and GRAVITY/GRAVITY+ a reality. 
~\\

\noindent {\bf Funding statement.}  
This project has received funding from the European Union's Horizon 2020 research and innovation programme under the Marie Sklodowska-Curie grant agreement No 101007855. 
This work was supported by Paris {\^I}le-de-France Region and by the French National Research Agency (ANR) under grant ANR-23-EDIR-0003 (GRAFITY).
The research leading to these results has received funding from the European Research Council (ERC) under the European Union's Horizon 2020 research and innovation program (project UniverScale, grant agreement 951549).
CC., NM and PG acknowledge support by FCT - Funda\c{c}\~ao para a Ci\^encia e a Tecnologia, I.P., Portugal, in the framework of the project Center for Astrophysics and Gravitation (CENTRA/IST/ULisboa) through grant No. UID/PRR/00099/2025 and grant No. 2022.01293.CEECIND/CP1733/CT0012.
    JC acknowledges financial support from ANID -- FONDECYT Regular 1251444, and Millennium Science Initiative Program NCN$2023\_002$.
    TPi acknowledges support from an advanced ERC grant MultiJets and from the Simons foundation SCEECS collaboration. The research of DC has been funded by the Alexander von Humboldt Foundation.
    JSt acknowledges funding from the Deutsche Forschungsgemeinschaft (DFG, German Research Foundation) under its Emmy-Noether Programm (STA 1955/1-1, Projektnummer 545534254).
  ~\\
    
\noindent {\bf Author contributions.}  
The GRAVITY+ Collaboration designed, built, commissioned and uses the VLTI-instrument GRAVITY+ for their observations. FMa, JO, SG analyzed the data presented here and prepared this manuscript. KAED, NA, AD, FE, AF, SG, XH, SJ, DL, SL, FMa, TO, TPa, CP, GP, DCR, DS, JSh, MSB, TTS, IU, FV, JW were acting as observers. RA, AB, JPB, GB, WB, CC, DD, FE, MF, HF, RGL, SG, XH, MH, JKo, RL, SL, OL, JBLB, BL, FMi, MM, NM, HN, SO, TO, KP, GP, RP, POP, SR, SRD, NP, JSB, JSa, SS, JSc, TTS, CSt, MS, PT, JW, GZ contributed to the soft- and hardware of GRAVITY(+). LD, FE, SFH, PK, SL, JL, MN, JO, FS, JSt, CSy  worked on the data reduction and the code for it. The project was managed by FE, LEO, PG, RG, SG, FG, JKa, LK, DL, AM, TO, TPa, ES, JW. The imaging of the data was done by AD, FM and JSt. RD, FE, NMFS, RG, AK, LK, LL, DL, TPa, KP, GP, CSt, LJT were critical in securing the project resources. GB, PAS, AB, DC, JC, FE, PG, AG, RG, SG, SJ, FMa, JO, TO, TPa, HBP, TPi, DCR, MSB, RS, FV worked on the interpretation and preparation of the manuscript.}

\end{document}